\documentclass[twocolumn]{article}
\usepackage{graphicx} 

\usepackage[labelfont=bf]{caption}

\usepackage[style=nature]{biblatex}

\usepackage{authblk}
\usepackage{amsmath} 
\usepackage{siunitx}
\usepackage{braket}
\usepackage{physics}
\usepackage{amssymb}
\usepackage{dsfont}
\usepackage[style=nature]{biblatex}

\usepackage{authblk}
\usepackage{amsmath} 
\usepackage{siunitx}

\usepackage{newfloat}
\DeclareFloatingEnvironment[name={Supplementary Figure}]{suppfigure}

\newcommand{\GS}{\SI{1110.755768\pm85.65643848}{\giga\hertz}}
\newcommand{\SpinSplitting}{\SI{254.654248\pm3.865435}{\mega\hertz}}
\newcommand{\FRandBench}{\SI{0.99809572042616 \pm 0.0004236464362647879}{}}
\newcommand{\FRandBenchTwo}{\SI{0.99485292\pm0.00453646}{}}
\newcommand{\cyclicity}{\SI{816.2852089805903\pm15.340142435396597}{}}
\newcommand{\epssimulated}{\SI{392}{\giga\hertz}}

\newcommand{\TimeTwoStar}{\SI{4.67294555\pm0.303319347}{\micro\second}}
\newcommand{\Aparallel}{\SI{621.75027\pm4.17829}{\kilo\hertz}}

\newcommand{\TimeTwoSLOffRes}{\SI{144.976015\pm10.5006019}{\micro\second}}
\newcommand{\TimeTwoSLResOne}{\SI{17.346\pm6.0912}{\micro\second}}

\newcommand{\AperpHH}{\SI{74.765\pm3.591}{\kilo\hertz}}

\newcommand{\TimeTwoCPMG}{\SI{273.0479434\pm14.52074}{\micro\second}}
\newcommand{\gammaCPMG}{\SI{0.40387594\pm0.009092620652459572}{}}

\newcommand{\TimeOne}{\SI{535.45\pm51.591125}{\micro\second}}
\newcommand{\TimeTwoStarNuclear}{\SI{0.927552\pm0.1899754}{\milli\second}}
\newcommand{\TimeTwoNuclear}{\SI{1.16688235\pm0.15763756}{\milli\second}}

\newcommand{\AperpXYfourtytwo}{\SI{140.1041\pm6.915190611593291}{\kilo\hertz}}
\newcommand{\OmegaLXYfourtytwo}{\SI{3.5857929\pm0.000682}{\mega\hertz}}
\newcommand{\FnucXYfourtytwoSim}{\SI{0.647}{}}
\newcommand{\FnucXYfourtytwoRam}{\SI{0.681813413771746 \pm 0.0483536917713895}{}}
\newcommand{\taunucinit}{\SI{81.5}{\nano\second}}
\newcommand{\electrong}{\SI{2.0123970845673194\pm0.00038297868664406904}{}}

\newcommand{\OmegaRabiNucCond}{\SI{42.3}{\kilo\hertz}}
\newcommand{\OmegaRabiNucUncond}{\SI{5.7}{\kilo\hertz}}

\newcommand{\AparallelNuclearRamsey}{\SI{572.9741477859767 \pm 53.4022379198438}{\kilo\hertz}}

\newcommand{\LowRabi}{\SI{18.911032332912888\pm0.7197166647441085}{\kilo\hertz}}
\newcommand{\LowGammaTwoR}{\SI{1.4245884263240187\pm0.1023229522555123}{\kilo\hertz}}

\newcommand{\TimeTwoRabiCNOT}{\SI{64.99466526347477\pm3.491858368750885}{\micro\second}}
\newcommand{\TimeTwoPiCNOT}{\SI{2.8706505831223947\pm0.003997685394125108}{\micro\second}}

\newcommand{\FInitBright}{\SI{0.935242183637534 \pm 0.03178135617273691}{}}
\newcommand{\FInitDark}{\SI{0.827308120854815 \pm 0.0452152149308865}{}}

\newcommand{\PumptimeNuclear}{\SI{41.6178057\pm 0.642350676}{\milli\second}}
\newcommand{\PumptimeNuclearRev}{\SI{39.9225263 \pm 0.56246321}{\milli\second}}
\newcommand{\PumpContrastNuclear}{\SI{0.6813482522537093 \pm 0.0012563666813520662}{}}
\newcommand{\PumpContrastNuclearRev}{\SI{0.6728560232537629 \pm 0.001194842594059777}{}}

\newcommand{\FourierLimit}{\SI{96.2510\pm21.965155350272397}{\mega\hertz}}
\newcommand{\OmegaROptMax}{\SI{1.14422658\pm0.02546047987776062}{\giga\hertz}}
\newcommand{\TimeTwoOpt}{\SI{1.18183720\pm0.0892911248}{\nano\second}}

\newcommand{\btheta}{\SI{28}{\degree}}
\newcommand{\alphasimulated}{\SI{0.68}{}}
\newcommand{\wLesimulated}{\SI{9.437}{\giga\hertz}}
\newcommand{\etasimulated}{\SI{816}{}}
\newcommand{\gsssimulated}{\SI{1111}{\giga\hertz}}
\newcommand{\dsssimulated}{\SI{255}{\mega\hertz}}

\newcommand{\LinewidthLeft}{\SI{114.98043\pm6.801982199}{}}
\newcommand{\LinewidthRight}{\SI{113.23165734\pm7.4093840}{\mega\hertz}}
\newcommand{\SplittingAverage}{\SI{254.654248\pm3.865435}{\mega\hertz}}
\newcommand{\Groundstatesplitting}{\SI{1110.7548797379573\pm85.65643898050575}{\giga\hertz}}
\newcommand{\wLe}{\SI{9.431}{\giga\hertz}}
\newcommand{\Cyclicity}{\SI{816.2852095\pm15.34014243}{}}

\newcommand{\RabiHH}{\SI{3.597139\pm0.002865}{\mega\hertz}}

\newcommand{\AperpHHParabola}{\SI{74.3075040\pm1.755275083107678}{\kilo\hertz}}

\newcommand{\alphaone}{\SI{2.17055301 \pm 0.245996911}{}}
\newcommand{\alphatwo}{\SI{1.54312249 \pm 0.0629408370}{}}
\newcommand{\bttwostar}{\SI{3.39683957 \pm 1.48512155}{}}

\newcommand{\wLXYtwentyfour}{\SI{3.5849685\pm0.00152196}{\mega\hertz}}
\newcommand{\AperpOneXYtwentyfour}{\SI{163.497\pm13.16}{\kilo\hertz}}
\newcommand{\AperpTwoXYtwentyfour}{\SI{70.187536\pm36.128874}{\kilo\hertz}}

\newcommand{\wLXYthritytwo}{\SI{3.584564073\pm0.000586}{\mega\hertz}}
\newcommand{\AperpOneXYthritytwo}{\SI{141.200054\pm8.163816081937744}{\kilo\hertz}}
\newcommand{\AperpTwoXYthritytwo}{\SI{104.477953\pm12.874904}{\kilo\hertz}}

\newcommand{\wLXYfourtytwo}{\SI{3.5857929\pm0.000682}{\mega\hertz}}
\newcommand{\AperpOneXYfourtytwo}{\SI{140.1041\pm6.915190611593291}{\kilo\hertz}}
\newcommand{\AperpTwoXYfourtytwo}{\SI{101.19309\pm10.865998}{\kilo\hertz}}

\newcommand{\wLXYsixtyfour}{\SI{3.5841947\pm0.000539}{\mega\hertz}}
\newcommand{\AperpOneXYsixtyfour}{\SI{136.9368\pm7.36}{\kilo\hertz}}
\newcommand{\AperpTwoXYsixtyfour}{\SI{86.0056\pm3.02}{\kilo\hertz}}

\newcommand{\AperpOneXYInitRead}{\SI{136.828128\pm8.770472907004049}{\kilo\hertz}}
\newcommand{\AperpTwoXYInitRead}{\SI{106.27679\pm3.52877}{\kilo\hertz}}

\newcommand{\wLXYRcond}{\SI{3.582\pm0.003378189506718145}{\mega\hertz}}
\newcommand{\AperpOneXYRcond}{\SI{138.036685\pm2.933264593479157}{\kilo\hertz}}
\newcommand{\AperpTwoXYRcond}{\SI{107.7486\pm4.459637863756321}{\kilo\hertz}}

\newcommand{\wLXYRuncond}{\SI{3.58314081\pm0.002857441}{\mega\hertz}}
\newcommand{\AperpOneXYRuncond}{\SI{136.46195\pm2.7746224587807378}{\kilo\hertz}}
\newcommand{\AperpTwoXYRuncond}{\SI{110.48261\pm5.630357}{\kilo\hertz}}

\newcommand{\BetaNTimeTwoStar}{\SI{2.270535\pm1.452528}{}}
\newcommand{\BetaNTimeTwo}{\SI{1.75207734\pm0.5754832}{}}

\newcommand{\TimeOneOptical}{\SI{1.6535\pm0.37734892341174103}{\nano\second}}

\begin{document}
\title{Ultra-high strained diamond spin register with coherent optical link}
\author[1]{Marco Klotz$^*$}
\author[1]{Andreas Tangemann$^*$}
\author[1,2]{Alexander Kubanek$^\dagger$}
\affil[1]{Institute for Quantum Optics, Ulm University, Albert-Einstein-Allee 11, 89081 Ulm, Germany}
\affil[2]{Center for Integrated Quantum Science and Technology (IQST), Ulm University, Albert-Einstein-Allee 11, Ulm 89081, Germany}
\setcounter{Maxaffil}{0}
\renewcommand\Affilfont{\itshape\small}
\date{19th September 2024}

\twocolumn[
\begin{@twocolumnfalse}
	\maketitle
	\begin{abstract}
		Solid-state spin defects, such as color centers in diamond, are among the most promising candidates for scalable and integrated quantum technologies. 
		In particular, the good optical properties of silicon-vacancy centers in diamond combined with naturally occurring and exceptionally coherent nuclear spins serve as a building block for quantum networking applications. 
		Here, we show that leveraging an ultra-high strained silicon-vacancy center inside a nanodiamond allows us to coherently and efficiently control its electron spin, while mitigating phonon-induced dephasing at liquid helium temperature. 
		Moreover, we indirectly control and characterize a $^{13}\text{C}$ nuclear spin and establish a quantum register. 
		We overcome limited nuclear spin initialization by implementing single-shot nuclear spin readout. 
		Lastly, we demonstrate coherent optical control with GHz rates, thus connecting the register to the optical domain. 
		Our work paves the way for future integration of quantum network registers into conventional, well-established photonics and hybrid quantum communication systems. \vspace{0.5cm} 
	\end{abstract}
\end{@twocolumnfalse}
]
\def\thefootnote{*}\footnotetext{These authors contributed equally to this work}
\def\thefootnote{$\dagger$}\footnotetext{Corresponding author: alexander.kubanek@uni-ulm.de}
\section*{Introduction}
Color centers in diamond, due to their exceptional coherence properties, have shown to be promising candidates for a variety of quantum technologies such as quantum information processing and quantum networking applications \cite{atature2018material, ruf2021quantum, pompili2021realization, stas2022robust, knaut2024entanglement, awschalom2018quantum}. 
In particular, negatively-charged group-IV centers have recently attracted great attention due to their high Debye-Waller factor and inversion-symmetric optical dipole, which allows those defects to be highly stable even when integrated into and interfaced with nano-scaled photonic and phononic infrastructure \cite{sipahigil2016integrated, parker2024diamond, ngan2023quantum, clark2024nanoelectromechanical, kuruma2023engineering, antoniuk2024all, wan2020large}, thereby offering great promise with regards to scalability. 
However, electron spin coherence is limited due to phonon-induced dephasing in the ground-state orbitals at liquid-helium temperature. 
In order to suppress phonon interactions either their density of states or occupation has to be reduced. 
The former can be realized by tailoring the phononic environment \cite{kuruma2023engineering, klotz2022prolonged}, whereas the latter can be achieved by either cooling the system down to milli-Kelvin temperatures \cite{sukachev2017silicon, nguyen2019quantum} or increasing the ground-state splitting by using strained \cite{sohn2018controlling, stas2022robust, klotz2024strongly, assumpcao2023deterministic} or heavier defects \cite{senkalla2024germanium, rosenthal2023microwave, guo2023microwave, karapatzakis2024microwave}.
Here, we work with an ultra-high strained negatively-charged silicon vacancy center (SiV) with over \SI{1}{\tera\hertz} ground-state splitting inside a nanodiamond above \SI{2.6}{\kelvin}. 
This allows us to coherently control three different qubits, namely the SiV's electron spin, its optical dipole as well as a nearby, moderately-coupled $^{13}\text{C}$ nuclear spin. 
We first characterize the electron spin coherence properties reaching a dephasing time of $T_\mathrm{2,e}^{*} = \TimeTwoStar$, so far only achieved at 100mK \cite{stas2022robust}. 
Furthermore, we measure a coherence time $T_\mathrm{2,e} = \TimeTwoCPMG$ exceeding previous measurements with a less-strained SiV in a nano-photonic cavity \cite{stas2022robust, knaut2024entanglement} and compatible with recent reports from strained tin-vacancy centers (SnV) \cite{guo2023microwave} at our operating temperature. 
We proceed by dynamical-decoupling (DD) mediated nuclear spin control and characterization, leading to a $T_\mathrm{2,n}$ exceeding \SI{1}{\milli\second}, limited by electron spin relaxation. 
Continuously decoupling the electron spin enables us to overcome limited DD-mediated nuclear spin initialization by performing single-shot nuclear spin readout. 
Lastly, we also show coherent control of the SiV's optical dipole with GHz Rabi frequencies, thereby introducing a link between the two-qubit spin register and a photonic qubit. 
\section*{Results}
\subsubsection*{Electron spin control of an ultra-high strained SiV.}

\begin{figure*}[h]
	\centering
	\includegraphics[scale=1]{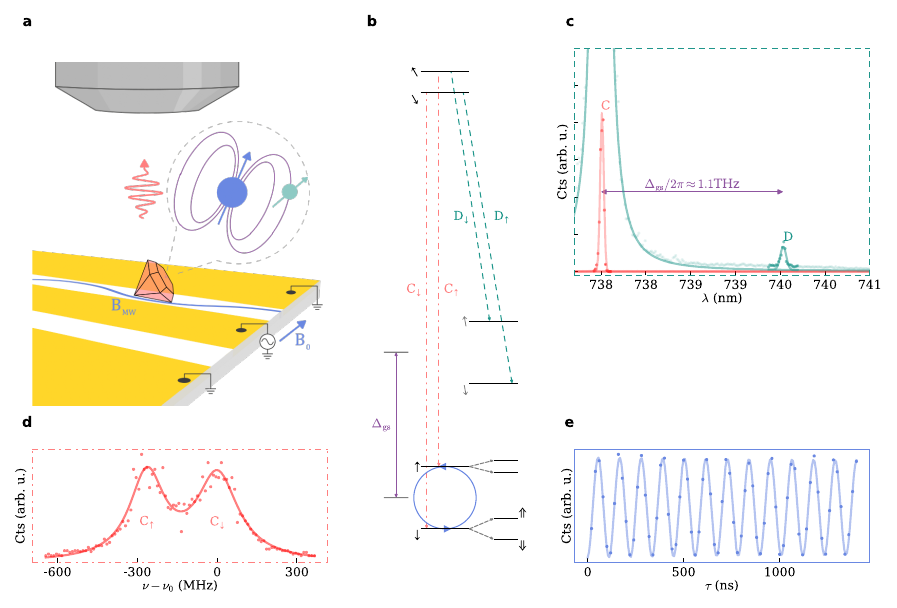}
	\caption{ \textbf{Electron spin control of an ultra-high strained SiV.} 
		\textbf{a} The nanodiamond containing the center is located on a sapphire substrate with a microwave antenna on-top delivering an alternating magnetic field $B_\mathrm{MW}$. 
		A static magnetic field $B_\mathrm{0}$ is supplied with permanent magnets. 
		\textbf{b} Energy diagram of the SiV. 
		Due to spin-orbit coupling and ultra-high strain, the SiV ground states are split by $\Delta_\mathrm{gs}/2\pi = \GS $ leading to the two transitions C and D. Higher energy transitions are omitted. 
		The Zeeman effect splits the degenerate spin-orbit states in electron spin states $\uparrow$ and $\downarrow$, which are further split due to hyperfine interaction with a nuclear spin, $\Uparrow$ and $\Downarrow$. 
		\textbf{c} Exciting the color center resonantly on transition C reveals transition D on a spectrometer (green). Red data is a highly-attenuated laser-reference signal. 
		\textbf{d} Photo-luminescence excitation spectroscopy of C while resonantly driving the electron spin with a microwave reveals spin-conserving transitions $C_\uparrow$ and $C_\downarrow$ split by $\Delta_\mathrm{ss}/2\pi=\SpinSplitting$.
		\textbf{e} Microwave-based coherent control of the SiV electron spin with Rabi frequency close to 10MHz.}
	\label{fig:Figure1}
\end{figure*}
Our system consists of a single SiV hosted in a nanodiamond placed in the gap of a coplanar microwave waveguide which is fabricated on sapphire, see Fig.\ref{fig:Figure1}a and Methods for further details.
The SiV is a point defect in the diamond lattice consisting of two carbon vacancies and an interstitial silicon atom with $D_\mathrm{3d}$ symmetry \cite{hepp2014electronic, gali2013ab, rogers2014electronic, thiering2018ab}. 
The inherent inversion symmetry makes the optical dipoles resilient to electric-field fluctuations \cite{sipahigil2014indistinguishable}, enabling integration into nano-structured hosts \cite{zuber2023shallow, waltrich2023two}. 
The orbital ground states are split by $\Delta_\mathrm{gs}$, illustrated in Fig.\ref{fig:Figure1}b and c, through spin-orbit interaction $\lambda/2\pi = \SI{50}{\giga\hertz}$ and static strain $\varepsilon$ \cite{meesala2018strain, sohn2018controlling, thiering2018ab}.
A static magnetic field $B_\mathrm{0}$ lifts the spin degeneracy and allows for optical pumping of the corresponding spin-cycling transitions, see Fig.\ref{fig:Figure1}d. 
The two lowest eigenstates, labeled $\uparrow$ and $\downarrow$, define the electron spin qubit. 

We are using an ultra-high strained ($\varepsilon \gg \lambda$) SiV with a ground-state splitting $\Delta_\mathrm{gs}/2\pi = \GS$, depicted in Fig.\ref{fig:Figure1}c, and operate at temperatures $T > \SI{2.6}{\kelvin} $, where phonon-induced spin-dephasing is expected to be highly suppressed \cite{meesala2018strain, sohn2018controlling, jahnke2015electron}. From numerical simulations we estimate a strain magnitude of $\varepsilon/2\pi\approx \epssimulated$, see Supplementary Information \cite{SI}. 
In this regime the orbital and spin part of the qubit states are strongly decoupled \cite{Nguyen2019}. 
Consequently, efficient microwave driving while introducing minimal heat load is possible. 
The qubit is optically initialized with a maximum fidelity of $F_\mathrm{I,e}\approx 0.84 $ and a cyclicity $\eta \approx \cyclicity$ \cite{SI}.
$F_\mathrm{I,e}$ is limited due to simultaneous depolarization by off-resonantly driving the other spin-cycling transition, split by $\Delta_\mathrm{ss}/2\pi = \SpinSplitting$, see Fig.\ref{fig:Figure1}d, in addition to spin-lattice relaxation, independently measured to be $T_{1,\mathrm{e}} = \TimeOne$ \cite{SI}. 
The qubit has a resonance frequency of $\omega_\mathrm{e}/2\pi \approx \SI{9.431}{\giga\hertz} $ and we can coherently drive it with a Rabi frequency $\Omega_\mathrm{R}/2\pi $ up to $\SI{10}{\mega\hertz}$ shown in Fig.\ref{fig:Figure1}e, with a single-qubit gate fidelity of $F_\mathrm{G} = \FRandBenchTwo$, extracted from a Randomized Benchmarking experiment \cite{SI}. 
This is exceeding recent measured gate fidelities for SnV at similar temperatures \cite{rosenthal2023microwave, guo2023microwave}, which is a direct consequence of the ultra-high strain. 

\subsubsection*{Electron spin characterization.} 
\begin{figure*}
	\centering
	\includegraphics[scale=1]{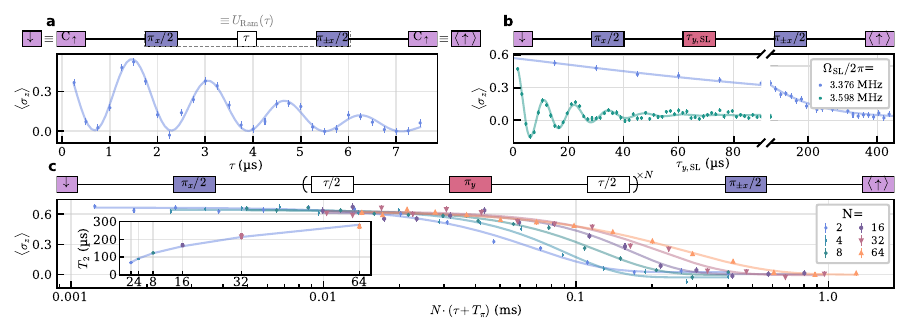}
	\caption{\textbf{Electron spin characterization.}
		\textbf{a} Electron Ramsey interference measurement shows coupling to a nearby spin with $A_{\parallel} /2\pi = \Aparallel $ and a dephasing time of $T_\mathrm{2,e}^{*} = \TimeTwoStar$. 
		Solid line is a fit to $(a_\mathrm{sin} \sin(A_{\parallel}\tau+\phi)+a_\mathrm{exp})\exp(-(\tau/T_\mathrm{2,e}^{*})^\beta) + c$. 
		\textbf{b} Continuously driving with frequency $\Omega_\mathrm{SL}/2\pi$ locks the electron spin and extends the coherence time to $T_\mathrm{2,SL}= \TimeTwoSLOffRes$. 
		Blue solid line is a fit to $a_\mathrm{exp}\exp(-(\tau/T_\mathrm{2,SL})^\beta)$. 
		On resonance with the surrounding spins, electron-nuclear cross relaxation reduces $T_\mathrm{2,SL}$ to $\TimeTwoSLResOne$, green curve. 
		The oscillations of the electron coherence reveals a dominant coupling to a nuclear spin with $(A_{\perp}/2)/2\pi=\AperpHH$. 
		Solid line is a fit to $\sum_{i=1}^{2} \ a_i\cdot \sin( (A_{\perp,i}/2) \tau+\phi_{i}) \exp(-(\tau/T_\mathrm{2,SL,i})^\beta_{i}) + c$. 
		\textbf{c} Using CPMG-N sequences $T_\mathrm{2,e}$ is extended to \TimeTwoCPMG, extracted from fits of the form $a_\mathrm{exp}\exp(-(\tau/T_\mathrm{2,e})^{\beta}) + c$. 
		Inset shows scaling of $T_\mathrm{2,e} \propto N^{\gamma}$ with $\gamma = \gammaCPMG$}
	\label{fig:Figure2}
\end{figure*}
We proceed with a Ramsey interference experiment, shown in Fig.\ref{fig:Figure2}a, from which we obtain a dephasing time of $T_\mathrm{2,e}^{*} = \TimeTwoStar$, confirming that phonon-induced dephasing is no-longer a limiting factor for $T_\mathrm{2,e}^{*}$ as opposed to zero-strain \cite{rogers2014all, pingault2017coherent, harris2024coherence}. 
A similar value has been recently reported at \SI{100}{\milli\kelvin} \cite{stas2022robust}. 
Above \SI{1}{\kelvin} this is the highest $T_\mathrm{2,e}^{*}$ reported for any group-IV center in diamond so far \cite{stas2022robust, senkalla2024germanium, sohn2018controlling, guo2023microwave, rosenthal2023microwave, karapatzakis2024microwave}. 
The data also reveals coupling to a nearby spin with a parallel interaction strength of $A_\parallel /2\pi = \Aparallel$, which we attribute to a naturally occurring $^{13}\text{C}$ nuclear spin. 

We confirm coupling to a $^{13}\text{C}$ nuclear spin bath with Hartmann-Hahn double resonance measurements \cite{metsch2019initialization, hartmann1962nuclear} by spin-locking (SL) the electron spin with different $\Omega_\mathrm{SL}$, effectively dressing it and leading to cross-relaxation if $\Omega_\mathrm{SL}$ is close to the Larmor frequency $\omega_\mathrm{L,n}$ of the spin bath.  
Varying the spin-lock duration $\tau_\mathrm{SL}$ far away from resonance continuously decouples the spin by introducing an energy barrier of $\Omega_\mathrm{SL}$ in the dressed states, such that noise orthogonal to the spin-lock axis is suppressed extending $T_\mathrm{2,SL}$ to $\TimeTwoSLOffRes$ \cite{tan2013demonstration, bermudez2012robust, zalivako2023continuous, joos2022protecting}, see Fig.\ref{fig:Figure2}b. 
$T_\mathrm{2,SL}$ is limited by driving-induced heating, again highlighting the necessity for efficient driving. 
At resonance a dominant coherent coupling with $(A_{\perp}/2)/2\pi=\AperpHH$ to a nuclear spin reduces the coherence to $T_\mathrm{2,SL} =\TimeTwoSLResOne$ \cite{metsch2019initialization} \cite{SI}. 
Moreover, the data reveals the presence of additional harmonics in the signal, most likely due to further nuclear spins.

We extend the coherence time $T_\mathrm{2,e}$ further by performing CPMG-N pulsed DD to $\TimeTwoCPMG$ with N=64 $\pi$ pulses. 
We observe exponential stretching factors $\beta$ varying between 1.5 and 2.5 and a scaling of $T_\mathrm{2,e} \propto N^\gamma$  with $ \gamma = \gammaCPMG$. We attribute the deviation from the expected scaling of $\gamma = 2/3$ for pure spin noise \cite{medford2012scaling, de2010universal, myers2014probing} to the onset of a remaining phonon-induced dephasing \cite{guo2023microwave}. 
However, this coherence time still exceeds $T_\mathrm{2,e}$ measured at $\SI{4}{\kelvin}$ for less-strained SiV \cite{stas2022robust} and is within the same order of magnitude as recent reports from heavier defects like SnV \cite{guo2023microwave}.

\subsubsection*{Electron-mediated nuclear spin control and characterization.} 
\begin{figure*}
	\centering
	\includegraphics[scale=1]{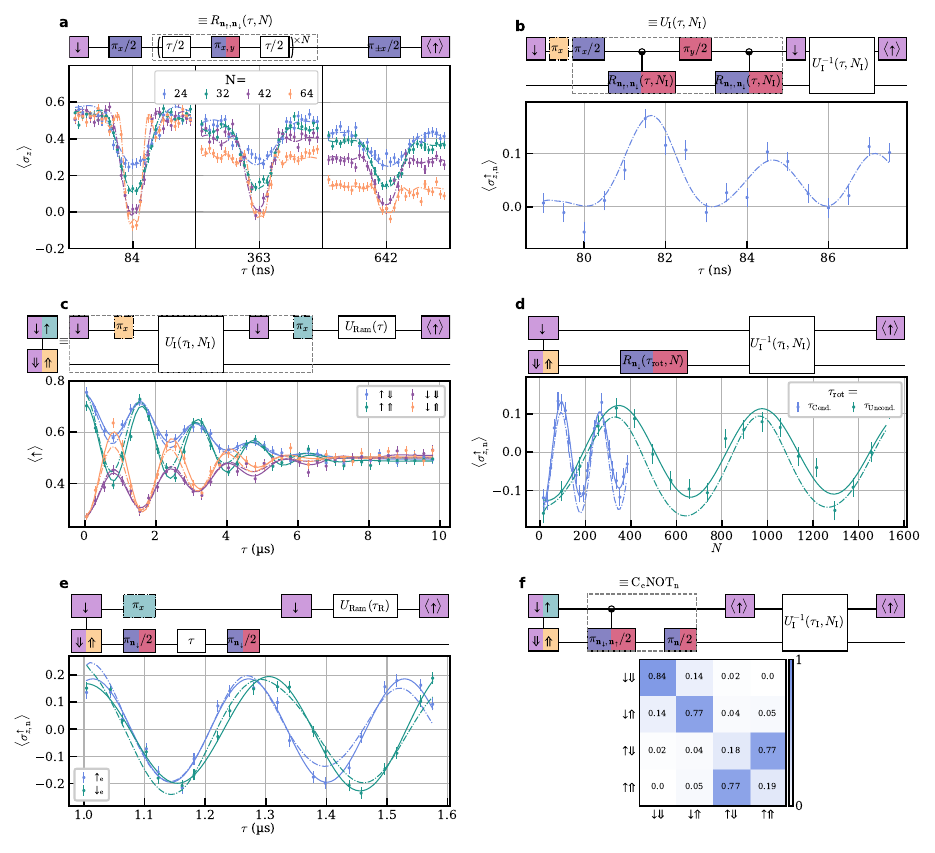}
	\caption{\textbf{
			Electron-mediated nuclear spin control and characterization.} 
		Solid lines are fits with phenomenological functions. 
		Dashed-dotted lines are obtained with a model with parameters listed in \cite{SI}. 
		\textbf{a} XY-N DD leads to periodic loss of coherence due to entanglement with the nuclear spin bath. 
		\textbf{b} The sequence $U_{\mathrm{I}}(\tau, N)$ transfers the population of the electron to the target nuclear spin, depending on $\tau$ and $N$ \cite{Nguyen2019, taminiau2014universal}. 
		$\tau$ is swept around the first resonance from \textbf{a} and $N_\mathrm{I}=42$. 
		Reversing $U_{\mathrm{I}}(\tau, N)$ followed by an electron spin readout probes the nuclear polarization. An additional $\pi$ pulse initializes the opposite nuclear spin, indicated with a yellow dash-dotted block. 
		\textbf{c} Ramsey interference after preparing all four basis states of the electron-nuclear two-qubit system. Data is fit to $(a_\mathrm{sin} \sin(A_{\parallel}\tau+\phi)+a_\mathrm{exp})\exp(-(\tau/T_\mathrm{2,e}^{*})^\beta) + c$. 
		\textbf{d} An increase of the number of $\pi$ pulses $N$ during DD coherently rotates the nuclear spin around an un-/conditioned rotation axis depending on $\tau_\mathrm{rot}$. 
		\textbf{e} Electron spin dependent Ramsey interference on the nuclear spin shows different precession frequencies due to $A_\parallel$. 
		\textbf{f} Amplitude transfer matrix of a $\mathrm{C_eNOT_n}$, composed of a conditional and unconditional $\pi/2$ rotation of the nuclear spin.}
	\label{fig:Figure3}
\end{figure*}
Using XY-N type of pulsed DD effectively applies a rotation $R_{\mathbf{n}_\uparrow,\mathbf{n}_\downarrow}(\tau, N)$ to the nuclear spins, illustrated in Fig.\ref{fig:Figure3}a, around an axis $\mathbf{n}$ which depends on the inter-pulse spacing $\tau$ and the electron spin's initial state $\uparrow$, $\downarrow$ \cite{Nguyen2019, maity2022mechanical, takou2023precise}. 
The rotation angle depends on $\tau$ and the number $N$ of $\pi_{x/y}$ pulses \cite{maity2022mechanical}. 

The conditional rotation is clearly visible in periodic resonances in Fig.\ref{fig:Figure3}a, where nuclear spins start to entangle with the electron spin, thereby reducing its coherence \cite{Nguyen2019, zahedian2024blueprint}. 
From the fit of a numerical model applied to the XY-$42$ measurement \cite{maity2022mechanical}, we obtain a perpendicular hyperfine parameter and Larmor frequency of the target nuclear spin, $A_\perp /2\pi = \AperpXYfourtytwo$ and $\omega_\mathrm{L,n}/2\pi = \OmegaLXYfourtytwo$, inline with previous Hartmann-Hahn measurements. 
See Methods and \cite{SI} for further details of the model and fit values.
Moreover, taking the gyromagnetic ratio $\gamma_{^{13}C}$ of $^{13}C$ and the implied magnetic field $B_0=\omega_\mathrm{L,n}/\gamma_{^{13}C}$ leads to an electron g-factor of $g_\mathrm{e}=\electrong$, close to that of a free electron, indicating the decoupled orbit and spin degree of freedom.

Following ref. \cite{Nguyen2019, maity2022mechanical} we use rotations on the electron spin in conjunction with electron-mediated nuclear rotations to construct an initialization gate $U_\mathrm{I}(\tau, N)$ which transfers the electron spin's population onto the target nuclear spin followed by re-initialization of the electron spin. 
Applying a $\pi$ pulse before this gate initializes the opposite nuclear spin state. 
We subsequently reverse the gate and readout the electron spin to probe the nuclear spin population. Fig.\ref{fig:Figure3}b shows the measured data for $N_\mathrm{I} = 42$, where we infer a target nuclear spin initialization of $F_{\mathrm{I,n}} = \FnucXYfourtytwoSim$ at $\tau = \taunucinit $ with the numerical model \cite{SI}.

We also independently verify the transfer by performing a Ramsey measurement, see Fig.\ref{fig:Figure3}c, after initializing the nuclear and electron spin \cite{taminiau2014universal}. 
From the resonant and off-resonant oscillation amplitudes, $a_\mathrm{exp}$ and $a_\mathrm{osc}$, being proportional to the respective nuclear spin populations, we calculate an initialization fidelity $\langle F_\mathrm{I, \Uparrow/\Downarrow} \rangle = \langle a_\mathrm{exp/osc} / (a_\mathrm{exp} + a_\mathrm{osc}) \rangle =\FnucXYfourtytwoRam$, averaged over $\uparrow$ and $\downarrow$, in close agreement to the previous measurement in Fig.\ref{fig:Figure3}b.
We can directly infer the contrast from the target nuclear spin's polarization by fixing $\tau$ during Ramsey to $\tau_\mathrm{R} = \pi/A_\parallel$. 

Next, we also probe conditional and unconditional coherent nuclear spin rotations by selecting two different $\tau_\mathrm{rot} = T_\mathrm{L}/2-T_{\pi}$ and $T_\mathrm{L}-T_{\pi}$ \cite{taminiau2012detection}, where $T_\mathrm{L}$ is the Larmor period and $T_\pi$ the $\pi$-pulse duration and vary $N$ to increase the rotation angle \cite{taminiau2012detection, maity2022mechanical}, shown in Fig.\ref{fig:Figure3}d. 
We extract $N=169$ and $629$ for one full rotation, equaling a nuclear Rabi frequency of $\Omega_\mathrm{R,n} = \OmegaRabiNucCond$ and $\OmegaRabiNucUncond$, respectively. 
For the unconditional rotation we only have a moderate agreement to the model, which we attribute to coupling to additional weaker coupled nuclear spins and therefore uncertainties in the $A_{\perp}$ estimation \cite{SI}. 
It is worth noting that we can perform over $1\,\mathrm{k}$ $\pi$ pulses without significant loss of signal,  thanks to the high single-qubit gate fidelity. 

Continuing with nuclear spin coherence characterization, we perform an electron spin dependent nuclear Ramsey measurement, depicted in Fig.\ref{fig:Figure3}e, showing two distinct frequencies split by $\AparallelNuclearRamsey$ \cite{maity2022mechanical}, which agrees with $A_\parallel$ within error bounds.
Furthermore, we measure a $T^{*}_{2,\mathrm{n}} = \TimeTwoStarNuclear $ and with a single refocusing pulse a $T_{2,\mathrm{n}} = \TimeTwoNuclear$, inline with limitations through electron spin relaxation $T_{1,\mathrm{e}} = \TimeOne$. 
Data is presented in \cite{SI}.

Having determined the conditional and unconditional $\Omega_\mathrm{R,n}$, we can construct a $\mathrm{C_eNOT_n}$, i.e. a nuclear spin flip conditioned on the electron spin state, from a conditional and unconditional $\pi/2$ pulse. 
To this end, we again initialize the four electron and nuclear spin population combinations, apply the $\mathrm{C_eNOT_n}$ and readout both populations by inverting $U_\mathrm{I}$. 
We reference the readout to the same measurement with an identity operation instead of a $\mathrm{C_eNOT_n}$ \cite{Nguyen2019} and thereby infer an amplitude transfer matrix \cite{nguyen2019quantum}, shown in Fig.\ref{fig:Figure3}f. 
The discrepancy to an optimal $\mathrm{C_eNOT_n}$ can be explained by reconsidering the electron initialization fidelity of $F_\mathrm{I,e}\approx0.84$, indicating an actual $\mathrm{C_eNOT_n}$ gate fidelity of well above $0.9$.

\subsubsection*{Nuclear-spin controlled electron flip and single-shot nuclear spin readout.} 

\begin{figure*}
	\centering
	\includegraphics[scale=1]{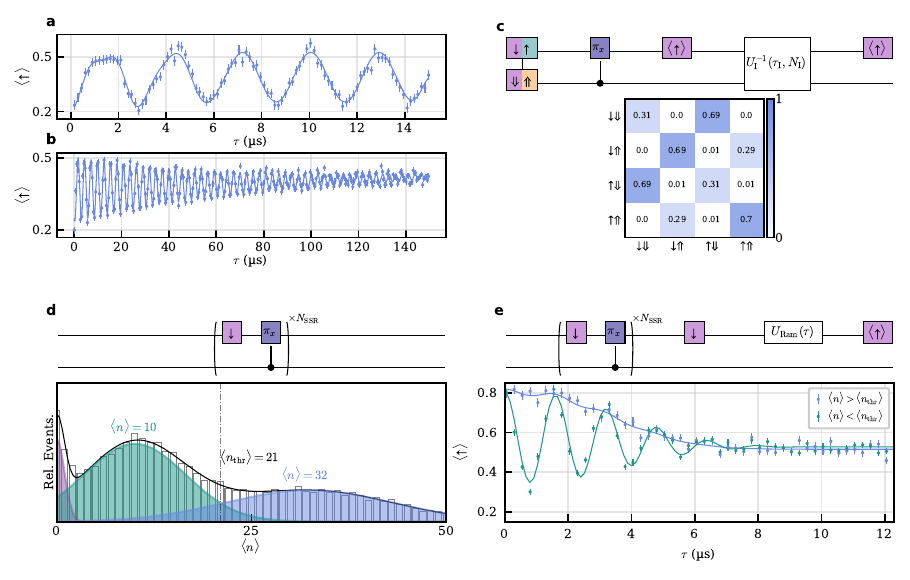}
	\caption{\textbf{Nuclear-spin controlled electron flip and single-shot nuclear spin readout.} 
		\textbf{a} Low-power driving of the hyperfine-split electron transition shows two Rabi frequencies. 
		The signal is fit with $\sum_{i =1}^{2}{} \ a_i\cdot \sin(\Omega_\mathrm{R,i} \tau+\phi_{i}) \exp(-(\tau/T_\mathrm{2,R,i})) + c$ (solid lines). 
		\textbf{b} Same measurement settings as \textbf{a} with longer driving times to extract $T_\mathrm{2,R}$. 
		\textbf{c} Amplitude transfer matrix of the $\mathrm{C_nNOT_e}$ gate. 
		\textbf{d} Mean-photon statistics for a $\SI{3}{\milli\second}$ single-shot readout (SSR) by repeating a laser pulse and $\mathrm{C_nNOT_e}$ $N_\mathrm{SSR}$ times. 
		The data is fit with a sum of three normal distributions. 
		The bright and dark states are clearly separated by mean-photon number $\langle n_\mathrm{b}\rangle=32$ and $\langle n_\mathrm{d} \rangle=10$. 
		\textbf{e} Ramsey interference measurement after SSR of the bright and dark nuclear spin state.}
	\label{fig:Figure4}
\end{figure*}
The efficient and continuous decoupling of the electron spin enables us to probe nuclear-split electron spin transitions by measuring Rabi oscillations down to $\Omega_\mathrm{R} = \LowRabi$ with a minimum decoherence rate of $\Gamma_\mathrm{2,R} = \LowGammaTwoR$, data is presented in \cite{SI}. 
Due to the spin 1/2 nature of SiV, this is not visible during pulsed DD \cite{Nguyen2019, zahedian2024blueprint}, given limited coherence. 

Next, we construct a $\mathrm{C_nNOT_e}$ gate by performing a conditional $\pi$ pulse on the electron spin through resonantly driving one transition at $\Omega_\mathrm{R, \Downarrow}/2\pi = 1/(\TimeTwoPiCNOT) \approx (A_{\parallel}/\sqrt{3})/2 \pi$, such that the off-resonant spin rotates at $\Omega_\mathrm{eff, \Uparrow} = \sqrt{\Omega_\mathrm{R, \Downarrow}^2 + A_\parallel^2} = 2 \Omega_\mathrm{R, \Downarrow}$. 
This results in a beating of the two Rabi frequencies visible in Fig.\ref{fig:Figure4}a. 
Moreover, for a driving frequency of $\Omega_\mathrm{R}$ we extract a $T_\mathrm{2,R} = \TimeTwoRabiCNOT $, indicating the high quality ($Q = T_\mathrm{2,R} / T_\pi \approx 40$) of the gate, see Fig.\ref{fig:Figure4}b. 

We initialize our two-qubit system and determine an amplitude transfer matrix for the $\mathrm{C_nNOT_e}$ gate, see Fig.\ref{fig:Figure4}c. 
Similar to the previous $\mathrm{C_eNOT_n}$ gate infidelity, the $\mathrm{C_nNOT_e}$ gate fidelity is limited by the initialization fidelity of the nuclear spin, also indicating a potential gate fidelity well above $0.9$.

We further utilize the $\mathrm{C_nNOT_e}$ gate to optically readout the nuclear spin state in a single shot (SSR) \cite{hesselmeier2024high, neumann2010single}. 
To this end, we apply a short laser pulse of length $T_\mathrm{p} \approx \SI{10}{\micro\second}$, where $T_\mathrm{p}$ is the electron spin polarization time, followed by a $\mathrm{C_nNOT_e}$ and repeat this $N_{\mathrm{SSR}}$ times. 
If the nuclear spin is in the control state, the electron spin can be recovered and the SiV re-excited again, resulting in discrete jumps of the emitted fluorescence. 
The mean-photon number $\langle n \rangle$ statistics for a $T_\mathrm{SSR} = \SI{3}{\milli\second}$ SSR reveals distinct bright and dark states with $\langle n_\mathrm{b} \rangle=32$ and $\langle n_\mathrm{d}\rangle=10$, shown in Fig.\ref{fig:Figure4}d. 
We attribute the distribution with $\langle n \rangle =0$ to states where the SiV is far off-resonant with our excitation laser \cite{neu2012photophysics, bradac2010observation}. 
We can discriminate the two nuclear spin states in post-processing by conditioning $\langle n \rangle$ during SSR on a threshold of $\langle n_\mathrm{thr}\rangle=21$ such that the initialization fidelity for the bright and dark state are equally optimized to $F'_\mathrm{I,n_b} = \SI{.925}{}$ and $F'_\mathrm{I,n_d} = \SI{.91}{}$, respectively.

We verify the initialization by again measuring Ramsey interference following a SSR and evaluating the respective oscillation amplitudes which results in $F_\mathrm{I, n_b}=\FInitBright$ and $F_\mathrm{I, n_d}= \FInitDark$, see Fig.\ref{fig:Figure4}e. 
$F_\mathrm{I, n_b}$ agrees well with $F'_\mathrm{I,n_b}$ and is limited by an on-set of optically induced nuclear spin polarization, data shown in \cite{SI}. 
We observe a loss of fluorescence when extending the SSR to $\approx \SI{140}{\milli\second}$, indicating a weak nuclear spin optical polarization within $T_{\mathrm{p,n}}=\PumptimeNuclear$. 
This can be explained by considering a different hyperfine coupling in the optically excited electron spin state and consequential nuclear spin mixing with respect to the ground-state such that the nuclear spin can undergo a spin-flip when its in the bright state \cite{hesselmeier2024high}. 
During the electron spin readout in the Ramsey measurement, we noticed a slight increase/decrease of the steady-state fluorescence for the dark/bright nuclear spin state, which we corrected for, see \cite{SI}.
We attribute this effect as well as the discrepancy of $F_\mathrm{I,n_d}$ and $F'_\mathrm{I,n_d}$ to our inability to classify the bright and dark nuclear spin state irrespective from the bright and dark SiV state. 

\subsubsection*{Coherent optical link.} 
\begin{figure*}
	\centering
	\includegraphics[scale=1]{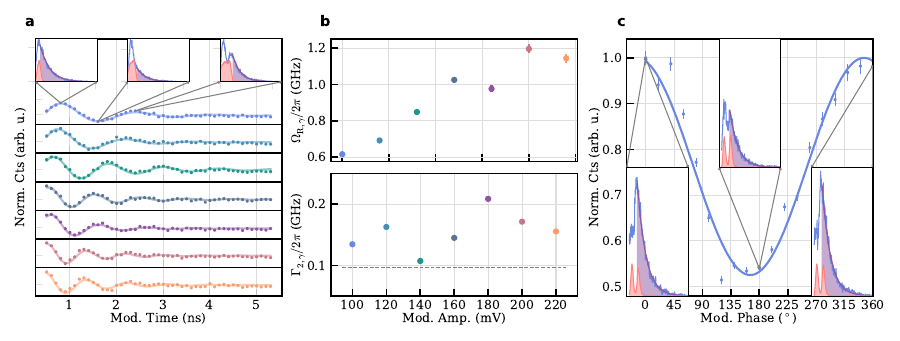}
	\caption{\textbf{Coherent optical link.} 
		\textbf{a} Rabi oscillations of the optical dipole for various driving strengths. 
		Insets show the measured fluorescence (blue), the excitation laser (red) and the fluorescence decay with corresponding fit (purple) which is proportional to the excited state population. 
		\textbf{b} Upper panel shows the Rabi frequency $\Omega_\mathrm{R, \gamma}$ as a function of the EOM's modulation amplitude. 
		Lower panel shows the decoherence rate $\Gamma_{2, \gamma}$. 
		Dashed line is the Fourier limit from independent lifetime measurements \cite{SI}. 
		Colours indicate the respective measurement from \textbf{a}. 
		\textbf{c} Varying the relative phase of two consecutive optical pulses enables complete control over the optical dipole. 
		Insets as in \textbf{a}. }
	\label{fig:Figure5}
\end{figure*}

In order to interconnect the two-qubit register a coherent optical link has to be established. We create short optical pulses by modulating the excitation laser’s amplitude with an electro-optical modulator (EOM). This creates sidebands whose duration, frequency, optical power and phase can be varied through the modulating microwave source. 
We use two etalons with a full-width-half-maximum (FWHM) of \SI{1.7}{\giga\hertz} to suppress the strong carrier. 
We choose the sidebands frequency to be in-between the two spin-dependent optical resonances such that we drive both transitions equally strong at various constant powers $(\Omega_\mathrm{R, \gamma}\gg \Delta_\mathrm{ss})$ while increasing the modulation time, see Fig.\ref{fig:Figure5}a. 
Extracting the subsequent exponential fluorescence decay, which is proportional to the optical excited-state population, shows coherent Rabi oscillations. 

Power-dependent measurements reveal the expected linear increase of $\Omega_\mathrm{R, \gamma}$ with the modulation amplitude, see Fig.\ref{fig:Figure5}b, with a maximum $\Omega_\mathrm{R, \gamma}/2\pi = \OmegaROptMax$. 
The optical decoherence rate $\Gamma_{2, \gamma}$ is consistently larger than the Fourier limit of $T_{1,\gamma}^{-1}/2\pi = \FourierLimit$ and tends to slightly increase with excitation power, hinting to laser-induced decoherence. 
The Fourier limit is extracted from the statistics of the fluorescence decay constants $T_{1, \gamma}$, see SI. 

We extend the coherent control to an arbitrary axis by varying the relative phase of the driving laser. 
To this end, we apply an $\SI{0.35}{\nano\second}$ optical pulse at $\SI{90}{\milli\volt}$ modulation amplitude, followed by another identical but phase-shifted pulse, see Fig.\ref{fig:Figure5}c. 
The second pulse is preceded by a temporal buffer of $\SI{0.8}{\nano\second}$ to circumvent interference at the etalons. These parameters have been chosen to optimize the contrast.
Insets in Fig.\ref{fig:Figure5}c show the fluorescence and reference laser pulses, where the latter exhibits no signs of remaining phase-dependent interference. 
However, this buffer time is limiting the maximum contrast due to the relatively short decoherence time $T_{2,\gamma} = \TimeTwoOpt$. 
The population in the excited state after the second pulse shows an expected sinusoidal dependence on the relative phase, thus confirming complete coherent control over the SiV's optical dipole.  

\section*{Discussion}

In conclusion, we have shown coherent control of three different qubits, namely the electron spin and optical dipole of a single SiV and a $^\mathrm{13} \mathrm{C}$ nuclear spin in a nanodiamond.
The ultra-high strain allowed us to operate at elevated temperatures above \SI{2.6}{\kelvin} and to measure an electron spin dephasing time of \TimeTwoStar, not limited by phonons, enabling sensing of moderately-coupled $^\mathrm{13} \mathrm{C}$ nuclear spins. 
Using DD we extended the coherence time to \TimeTwoCPMG, the longest for SiV and comparable to recently reported SnV at liquid-helium temperatures. \\
Using SiVs with even larger $\Delta_\mathrm{gs}$ will further reduce phonon-induced electron spin decoherence and relaxation. 
Hence, addressing of additional $^\mathrm{13}\mathrm{C}$ nuclear spins via indirect electron spin control becomes possible. 
This will increase the nuclear spins' coherence times, so far limited by electron spin relaxation, increasing the number of accessible nuclear spins towards large-scale spin registers, as has been previously demonstrated with a nitrogen-vacancy center \cite{van2024mapping}.\\
Since the host of the SiV can be integrated with conventional photonics \cite{lettner2024controlling, bayer2023optical, schrinner2020integration} and plasmonics \cite{wu2022room}, a potentially higher photon-collection efficiency \cite{antoniadis2023cavity} enables electron spin single-shot readout \cite{rosenthal2024single, kindem2020control}, which circumvents the current limits of the initialization fidelity. 
Additionally, combining an ancillary nuclear spin with single-shot readout and a subsequent SWAP gate can also improve electron spin initialization.\\
Sub-ns optical-control pulses in conjunction with the aforementioned $T_\mathrm{2,e}^*$ can be used to generate a high number of time- or frequency-binned spin-photon entangled states, making this system a promising platform to explore various quantum communication protocols, such as memory-based quantum repeaters \cite{stas2022robust, pompili2021realization}, multi-photon cluster states \cite{lodahl2022deterministic, thomas2022efficient} and hybridization with opto-mechanical systems \cite{shandilya2021optomechanical, chen2024phonon}. Extending our system to more than one nuclear spin \cite{ruskuc2022nuclear} could further increase the complexity of such protocols.
Finally, it is worth mentioning that these results neither required a vector magnet nor a dilution refrigerator which drastically lowers technical overhead and thus positively affects scalability concerns.

\section*{Methods}
\subsubsection*{Sample and setup.}  
The synthesis of the nanodiamonds containing the SiV used in this work has been described elsewhere \cite{antoniuk2024all, lettner2024controlling, klotz2024strongly}.
We use a sapphire substrate due to its good thermal conductivity. 
For microwave supply we use optical lithography, electron-beam metal deposition and metal lift-off to fabricate a $\SI{200}{\nano\meter}$ thick gold coplanar waveguide on top of the sapphire substrate with $\SI{20}{\nano\meter}$ titanium adhesion layer in-between. 
The centre conductor and gap width, $\SI{23}{\micro\meter}$ and $\SI{10}{\micro\meter}$ respectively, have been chosen to match $\SI{50}{\ohm}$ impedance to reduce unwanted reflections. 
The nanodiamond examined in this work is inside the $\SI{10}{\micro\meter}$ gap, where maximum microwave coupling is expected. 
The sapphire is placed on a custom copper coldfinger in a helium continuous-flow cryostat (Janis ST-500). 
A home-built confocal microscope with a room temperature 0.95-NA objective (Olympus MPLAPON) is used to excite the SiV and collect its fluorescence. 
Resonant excitation is done with a tunable Ti:Sapphire laser (Sirah). 
The emitted fluorescence from the phonon sideband of the SiV is filtered with a bandpass (Semrock 769/41) and detected with a superconducting nanowire single-photon detector (PhotonSpot) and subsequently time tagged with a TimeTagger Ultra (Swabian Instruments). The whole experiment is controlled with Qudi \cite{binder2017qudi}.
The in-plane static magnetic field is supplied with four standard neodymium permanent magnets (N52) in a Hallbach configuration buried in the cold finger underneath the sample. 
Microwave pulses are synthesized using a \SI{65}{\giga S \per \second} Arbitrary Waveform Generator (AWG) (Keysight M8195A), subsequently amplified (Minicircuits ZVE-3W-183+), sent to the cryostat and finally terminated outside the cryostat to reduce heatload.
Resonant optical pulses for readout and initialization are generated with an acousto-optical modulator (G\&H 3350-199), controlled by the AWG. 
For fast coherent optical-control pulses we use the laser-sidebands created with an amplitude electro-optical modulator (JENOPTIK AM705) which is locked with a PID and lock-in amplifier (Toptica DLC Pro) to half its maximum transmission and modulated with amplified (Minicircuits ZVE-3W-183+) \SI{7}{\giga \hertz} microwave pulses of the AWG. The laser carrier is blocked by two etalons (Laseroptik) with a FWHM of \SI{1.7}{\giga \hertz} and \SI{20}{\giga \hertz} free spectral range.
\newline

\subsubsection*{Spin initialization and readout}
We use a laser pulse resonant with the transition $C_\uparrow$ to initialize the electron spin in the state $\ket{\downarrow}$ and to read out $\ket{\uparrow}$. We determine the initialization fidelity $F_\mathrm{I,e}$ before every sequence by applying a microwave $\pi$ pulse after pumping the spin to the state $\ket{\downarrow}$ and reading it out with a second resonant laser pulse. We fit an exponential decay $a_\mathrm{exp}\exp(-\tau/T_\mathrm{p}) + n_\mathrm{ss}$, with polarization time $T_\mathrm{p}$ and steady-state counts $n_\mathrm{ss}$, to the fluorescence counts of the readout laser and calculate the initialization fidelity as $F_\mathrm{I,e}=a_\mathrm{exp}/(a_\mathrm{exp}+n_\mathrm{ss})$. We then extract the spin populations during the sequence by fitting the same function to each readout laser, where $T_\mathrm{p}$ and $n_\mathrm{ss}$ are bounded to values extracted from the collective signal. Details can be viewed in the SI.

\subsubsection*{System model and numerical simulation.}
We are using a model implemented with QuTIP \cite{qutip1, qutip2} to model the dynamics of our spin system. In a frame rotating with the microwave (MW) frequency $\omega_\mathrm{MW} / 2\pi$, the Hamiltonian of our system takes the form $(\hbar=1)$:
\begin{align}
\hat{H} = 
&\frac{\Delta}{2} \hat{\sigma}^\mathrm{e}_{z} + \frac{\Omega_\mathrm{R}}{2} \left( \cos \phi \hat{\sigma}^\mathrm{e}_{x} + \sin \phi \hat{\sigma}^\mathrm{e}_{y}\right) + \frac{\omega_\mathrm{L,n}}{2}\hat{\sigma}^\mathrm{n}_{z,i} + \nonumber \\
&\sum_{i=1}^{2} \hat{\sigma}^\mathrm{e}_{z} \left( \frac{A_{\parallel,i}}{4} \hat{\sigma}^\mathrm{n}_{z,i} + \frac{A_{\perp,i}}{4}\hat{\sigma}^\mathrm{n}_{x,i}\right)   \; ,
\end{align}
with the detuning $\Delta = \omega_\mathrm{e} - \omega_\mathrm{MW}$, the MW amplitude and phase $\Omega_\mathrm{R}$ and $\phi$, hyperfine-coupling strengths $A_{\parallel,i}$ and $A_{\perp,i}$ and the nuclear Larmor frequency $\omega_\mathrm{L,n}$. $\hat{\sigma}_i$ are Pauli operators.
We are using one target nuclear spin and a second one to account for infidelity due to interaction with the nuclear spin bath. 
We further use decoherence-free rotations of the electron spin of the form $R = \exp \left( -i \hat{H} T \right)$, where $T$ is the duration of the MW pulse. 
To account for loss of coherence during free evolution we multiply off-diagonal terms of the electron spin's density matrix with an empirical function $\exp{(-t/t_c)^\beta}$. We leave $t_c$ and $\beta$ as free parameters when fitting data to our model. 
\section*{Data Availability}
The data that support the findings of this study are available from the corresponding author upon reasonable request.
\section*{Acknowledgements}
We thank V.A. Davydov and V. Agafonov for their initial contribution to the nanodiamond material.
We thank Martin Plenio and Qiong Chen for fruitful discussion. 
The project was funded by the German Federal Ministry of Research (BMBF) in the project HybridQToken (16KISQ043K). We gratefully acknowledge further funding by the BMBF via the projects SPINNING (13N16215) and QR.X (16KISQ006) and by the European Union Program QuantERA in the project SensExtreme (499192368).

\section*{Author information}
\subsection*{Contributions}
AK supervised the project. MK and AT conceived the experiments. AT fabricated the sample. MK and AT set-up and conducted the experiments. MK designed the theoretical models and analyzed the data together with AT. MK, AT and AK discussed the data and wrote the manuscript. MK und AT contributed equally to this work.
\subsection*{Corresponding author}
Correspondence: alexander.kubanek@uni-ulm.de
\section*{Competing interests}
The authors declare no competing interests.

\setcounter{figure}{0}
\onecolumn
\subsection*{Supplementary Note 1: System parameter estimation.}
Closely following results from previous works on simulation of group-IV defects \cite{thiering2018ab, rosenthal2023microwave}, we here use QuTIP \cite{qutip1,qutip2} to numerically simulate ($\hbar=1$)
\begin{align*}
\hat{H} = \hat{H}_\mathrm{C} + \sum_{i=\mathrm{g,u}}\hat{H}^{i}_\mathrm{SO} + \hat{H}^{i}_\mathrm{Z, S} + \hat{H}^{i}_\mathrm{Z, L}  + \hat{H}^{i}_\mathrm{Str}
\end{align*}
and extract and estimate various system parameters. g/u stand for gerade/ungerade and describe the parity of orbital states. \\
Here 
\begin{align*}
\hat{H}_\mathrm{C}= \frac{\omega_\mathrm{C}}{2} \hat{\sigma}_z \otimes \mathds{1} \otimes \mathds{1} 
\end{align*}
is a Coulomb Hamiltonian which splits states with gerade/ungerade symmetry. We choose $\omega_C$ such that the C transition has \SI{736.9}{\nano\meter}. The first operator, $\hat{\sigma}_z = \ket{\mathrm{u}}\bra{\mathrm{u}} - \ket{\mathrm{g}}\bra{\mathrm{g}}$, acts on the parity of the states. The second operator acts in the subspace of the orbital states $\ket{e_x}, \ket{e_y}$ and the last operator acts in the spin subspace $\ket{\downarrow}, \ket{\uparrow}$. \\
The spin-orbit Hamiltonian takes the form
\begin{align*}
\hat{H}^\mathrm{g/u}_\mathrm{SO}= -\frac{\lambda^\mathrm{g/u}}{2} \; \hat{\sigma}_\mathrm{g/u} \otimes (-\hat{\sigma}_y) \otimes \hat{\sigma}_z \;, 
\end{align*}
where $\hat{\sigma}_\mathrm{g/u} = \ket{\mathrm{g/u}}\bra{\mathrm{g/u}}$ projects into the gerade/ungerade subspace. $\hat{\sigma}_y = -i \ket{e_x}\bra{e_y} + i \ket{e_y}\bra{e_x}$ is a Pauli matrix and used to describe the orbital momentum operator $\hat{L}_z = -\hat{\sigma}_y$. $\lambda^\mathrm{g/u}$ describes the spin-orbit interaction strengths with $2\pi \cdot 50/260$ GHz depending on the parity of the orbital states, respectively \cite{thiering2018ab}.\\
An applied magnetic field $B$ further introduces energy contributions through the Zeeman effect. There is an orbital Zeeman Hamiltonian which only contains a contribution along the symmetry z-axis (due to the defects D$_{\mathrm{3d}}$ symmetry),
\begin{align*}
H^\mathrm{g/u}_\mathrm{Z,L}&= \mu_\mathrm{B} \; p^\mathrm{g/u} \;  g_\mathrm{L}^\mathrm{g/u} \; B_z \; \underbrace{\hat{\sigma}_\mathrm{g/u} \otimes (-\hat{\sigma}_y) \otimes \mathds{1}}_{\hat{L}^\mathrm{g/u}_z} \; .
\end{align*}
$\mu_\mathrm{B}$ is the electron's Bohr magneton and $p^\mathrm{g/u}=0.308/0.128$, $g_\mathrm{L}^\mathrm{g/u}=0.328/0.782$ are parameters taken from \cite{thiering2018ab}.\\
There is also a spin Zeeman Hamiltonian
\begin{align*}
\hat{H}^\mathrm{g/u}_\mathrm{Z,S}&= \mu_\mathrm{B} \; g_\mathrm{S} \; \mathbf{\hat{S}}\cdot\mathbf{B} + \mu_\mathrm{B}2\delta_\mathrm{p}^\mathrm{g/u}\;g_\mathrm{L}^\mathrm{g/u} \;\hat{S}_z \; B_z \; ,
\end{align*}
where $g_\mathrm{S}=2.0023$ and which also contains an anisotropy term proportional to $\delta_\mathrm{p}^\mathrm{g/u}=0.003/0.028$. Again, parameters are taken from  \cite{thiering2018ab}. \\
The last term consists of strain contributions 
\begin{align*}
\hat{H}^\mathrm{g/u}_\mathrm{Str}&= \hat{\sigma}_\mathrm{g/u} \otimes \left( \varepsilon^\mathrm{g/u}_x  \; \hat{\sigma}_z + \varepsilon^\mathrm{g/u}_y  \; \hat{\sigma}_x\right) \otimes \mathds{1}
\end{align*}
with transverse strain parameters $\varepsilon^\mathrm{g/u}_x$ and $\varepsilon^\mathrm{g/u}_y$ and Pauli matrices $\hat{\sigma}_x, \hat{\sigma}_z$. We neglected axial strain contributions, since they only give rise to a common mode energy shift \cite{meesala2018strain, rosenthal2023microwave}. \\
We assume identical strain in $x$ and $y$ and parameterize $\varepsilon^\mathrm{g}_{x} = \varepsilon^\mathrm{g}_{y} = \varepsilon$ and $\varepsilon^\mathrm{u}_{x/y} = \alpha \varepsilon^\mathrm{g}_{x/y}$ , effectively also assuming identical strain susceptibility ratios in the gerade/ungerade subspace for $x$ and $y$. \\
We then calculate all eigenenergies $\mathrm{E}_{j=0 \dots 7}$ and eigenstates $\ket{\mathrm{e}_{j=0 \dots 7}}$ of the above Hamiltonian as a function of the magnetic field amplitude $B$, its angle to the symmetry axis $B_\theta$, strain magnitude $\varepsilon$ and $\alpha$. \\
We fix $B=\SI{335}{\milli\tesla}$, which we extracted from the $^{13}\mathrm{C}$ nuclear Larmor frequency and gyromagnetic ratio from independent Hartmann-Hahn and XY-N type measurements. Moreover, we assume $B_\phi=0$. \\
From the eigenenergies and eigenstates, we can extract experimentally accessible parameters, such as the electron spin resonance frequency
\begin{align*}
\omega_\mathrm{L,e} / 2\pi = \mathrm{E}_1 - \mathrm{E}_0,
\end{align*}
the splitting between the spin-dependent optical transitions $C_\uparrow - C_\downarrow$
\begin{align*}
\Delta_\mathrm{ss} / 2\pi = \left(\mathrm{E}_5 - \mathrm{E}_1\right) - \left(\mathrm{E}_4 - \mathrm{E}_0\right),
\end{align*}
as well as the ground-state splitting from the difference in average energies by 
\begin{align*}
\Delta_\mathrm{gs} / 2\pi = \frac{1}{2}\left(\mathrm{E}_3 + \mathrm{E}_2\right) - \frac{1}{2}\left(\mathrm{E}_1 + \mathrm{E}_0\right).
\end{align*}
The cyclicity $\eta = \Gamma_\mathrm{cyc} / \Gamma_\mathrm{flip}$ is determined from the eigenstates by calculating the squared ratio of optical dipoles for a spin-preserving and spin-flipping transition, i.e.
\begin{align}
\eta = \frac{\sum_{i=x,y,z} \left( \bra{e_0} p_i \ket{e_4} \right)^2}{\sum_{i=x,y,z} \left( \bra{e_1} p_i \ket{e_4} \right)^2}, \label{eq:cyclcity}
\end{align}
where 
\begin{align*}
p_x &= \hat{\sigma}_x\otimes\hat{\sigma}_z\otimes\mathds{1},\\
p_y &= \hat{\sigma}_x\otimes-\hat{\sigma}_x\otimes\mathds{1},\\
p_z &= 2\; \hat{\sigma}_x\otimes\mathds{1}\otimes\mathds{1}
\end{align*}
are the orbital dipole operators resulting in optical transitions from ungerade to gerade subspaces with different polarizations \cite{rosenthal2023microwave, hepp2014electronic}.\\
In order to estimate $\varepsilon$, $\alpha$ and $B_\theta$ from our measured values of $\omega'_\mathrm{L,e} / 2\pi = \wLe$, $\Delta'_\mathrm{ss} / 2\pi = \SplittingAverage$, $\Delta'_\mathrm{gs} / 2\pi = \Groundstatesplitting$, $\eta' = \Cyclicity$, we calculate with the above model the respective parameters and then minimize a cost function of the form
\begin{align*}
{\sum_i \left(\left(p_i - p'_i \right)/p'_i\right)^2},
\end{align*}
where $p_i$ and $p'_i$ are the calculated and measured parameters.\\
With this procedure we obtain $\varepsilon / 2\pi\approx \epssimulated$, $\alpha \approx \alphasimulated$ and $B_\theta \approx \btheta$ resulting in 
\begin{equation*}
\begin{aligned}
\omega_{L,e} / 2\pi &\approx \wLesimulated\\
\Delta_\mathrm{ss} / 2\pi &\approx  \dsssimulated \\
\Delta_\mathrm{gs} / 2\pi &\approx \gsssimulated \\
\eta & \approx \etasimulated
\end{aligned}
\quad 
\begin{aligned}
\omega'_\mathrm{L,e} / 2\pi &= \wLe \\
\Delta'_\mathrm{ss} / 2\pi &= \SplittingAverage \\ 
\Delta'_\mathrm{gs} / 2\pi &= \Groundstatesplitting \\
\eta' &= \Cyclicity
\end{aligned}
\end{equation*}
which are in excellent agreement with our experimentally measured values.\\
In addition, we use the previously determined system parameters to calculate the spin-relaxation rate $\Gamma_1^\mathrm{2ph}$ from a two-phonon Orbach process, according to \cite{guo2023microwave, orbach1961spin} as a function of $B_\theta$ and $\Delta_\mathrm{gs}/2\pi$
\begin{align*}
\Gamma_1^\mathrm{2ph} \propto \frac{\Delta_\mathrm{gs}^3}{\mathrm{exp}\left( \Delta_\mathrm{gs} / k_\mathrm{B}T\right)-1}\frac{\left| d_{0,2} d_{1,2}^* + d_{0,3} d_{1,3}^*\right|^2}{\left(d_{0,2}^2 + d_{1,2}^2 + d_{0,3}^2 + d_{1,3}^2 \right)}, 
\end{align*}
where $d_{i,j} = \bra{\mathrm{e}_i} \hat{H}_\mathrm{Str,g} \ket{\mathrm{e}_j}$ are the respective phononic transition dipoles, $k_\mathrm{B}$ is Boltzmann's constant and $T=\SI{2.6}{K}$ the temperature. 

\begin{suppfigure}[h]
	\centering
	\includegraphics[width=1\linewidth]{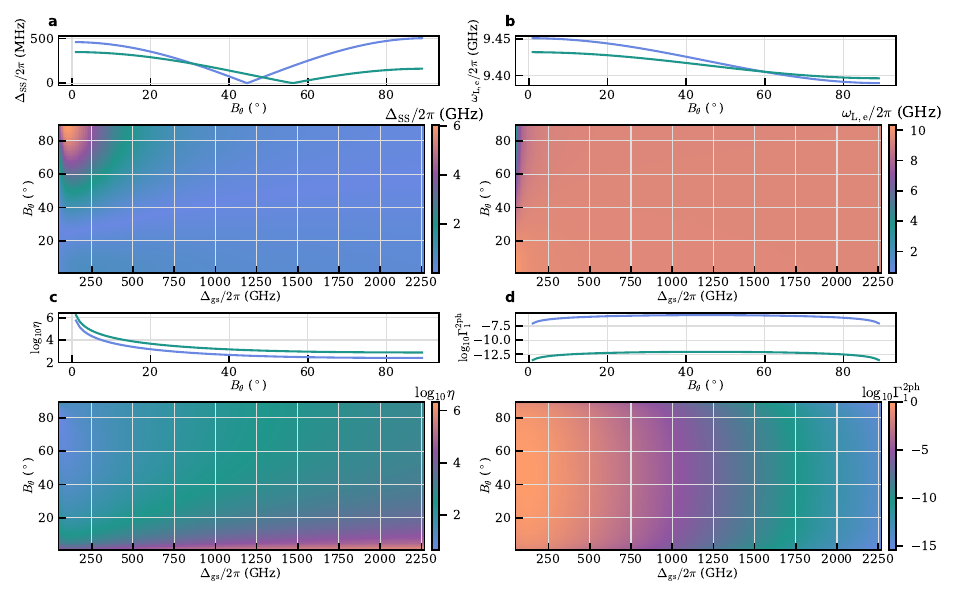}
	\caption{\textbf{System parameter estimation.} Simulated SiV system parameters for the previously estimated $\alpha \approx \alphasimulated$. \textbf{a} Spin splitting $\Delta_\mathrm{ss}/2\pi$, \textbf{b} electron spin Larmor frequency $\omega_\mathrm{L,e}/2\pi$, \textbf{c} cyclicity $\eta$ and \textbf{d} spin relaxation rate $\Gamma^\mathrm{2ph}_1$ by two phonon Orbach process as a function of ground-state splitting $\Delta_\mathrm{gs}/2\pi$ and magnetic field angle $B_\theta$ at a magnetic field strength $B = 335\,\mathrm{mT}$. Data is normalized to the maximum rate.
		Top panels shows data where $\Delta_\mathrm{gs}/2\pi = 1111/2000$ GHz for blue/green curve.}
	\label{fig:params}
\end{suppfigure}
First, from the comparison of $\Delta_\mathrm{gs}/2\pi = 1111$ GHz and $\Delta_\mathrm{gs}/2\pi = 2000$ GHz in the top panels of Supplementary Fig.\ref{fig:params}a we can see that an increased $\Delta_\mathrm{gs}/2\pi$ leads to a generally smaller separation of both spin-cycling transitions which would in turn impair spin initialization fidelity since off-resonant driving becomes stronger. 
Either applying a larger magnetic field or increasing its angle to the symmetry axis, $B_\theta$,  would improve the initialization fidelity. \\
Secondly, the trend in cyclicity $\eta$ is similar for both values of $\Delta_\mathrm{gs}/2\pi$, where a maximum $\eta$ is reached for $B_\theta=0$, i.e. maximal alignment. 
Additionally, there is a higher offset in $\eta$ for $\Delta_\mathrm{gs}/2\pi = 2000$ GHz due to a reduction in spin-mixing in the ground and excited state. 
This would facilitate potential electron spin single-shot readout, due to a longer polarization time before a spin flip occurs which allows for more photons to be collected. \\
Lastly, the two-phonon spin-relaxation rate $\Gamma_1^\mathrm{2ph}$ is also offset for larger strain with a similar functional dependence on $B_\theta$. 
Thus, for $\Delta_\mathrm{gs}/2\pi=2000$ GHz a significant increase in $T_1$ can be expected which would help improve the nuclear spin coherence, so far limited by $T_1$.
\newpage
\subsection*{Supplementary Note 2: Fidelity and data normalization.}

\begin{suppfigure}[h]
	\centering
	\includegraphics[width=1\linewidth]{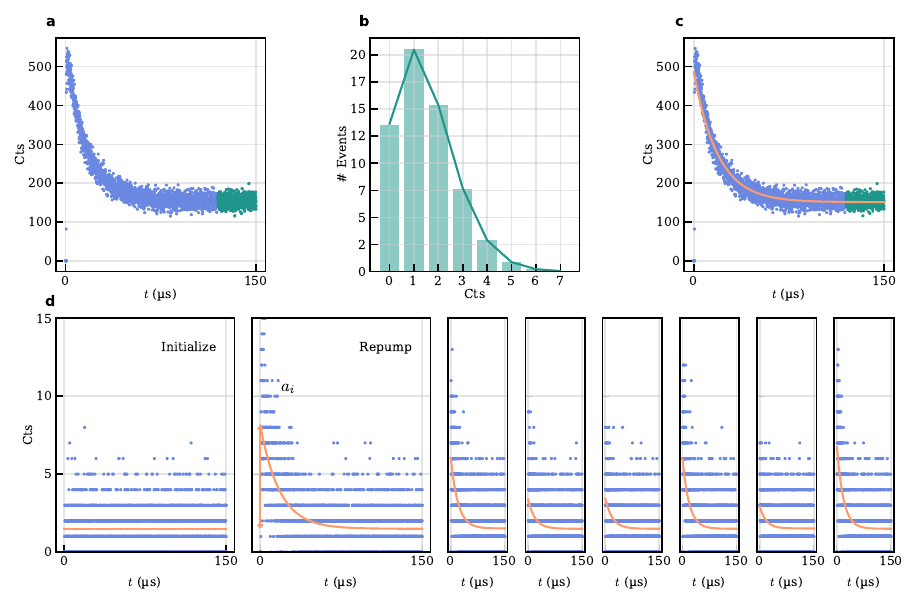}
	\caption{\textbf{Fidelity and data normalization.} 
		\textbf{a} Temporal histogram of the summed counts during one exemplary sequence. 
		\textbf{b} Histogram of the counts within the steady state, marked in green in \textbf{a}. 
		\textbf{c} Optical polarization time $T_\mathrm{p}$ of the summed laser pulses from \textbf{a}. 
		To extract $T_\mathrm{p}$ we fit an exponential decay, $a \exp(-t/T_\mathrm{p}) + c$. 
		\textbf{d} Amplitude extraction $a_i$ of each laser pulse with previously determined bounds for polarization time and steady state counts.}
	\label{fig:pulsefidelity}
\end{suppfigure}
We analyze the fluorescence during each laser pulse by first summing up all counts in each laser pulse, see Supplementary Fig.\ref{fig:pulsefidelity}a, to extract a mean-photon number $n_{\mathrm{ss}}$ within the steady state and a polarization time $T_\mathrm{p}$. 
To this end, we fit a Poisson distribution to the steady-state photon statistics, see Supplementary Fig.\ref{fig:pulsefidelity}b. 
Additionally, we fit a single exponential decay to the summed counts to infer the polarization time $T_\mathrm{p}$. 
Lastly, we take the counts in each laser pulse individually and again fit a single exponential decay where $T_\mathrm{p}$ and  $\langle n_{\mathrm{ss}} \rangle$ are now bounded to $\pm 3 \sigma$, where $\sigma$ is the standard deviation from the fit of Supplementary Fig.\ref{fig:pulsefidelity}a and b such that we can extract each individual amplitude $a_\mathrm{i}$. 
Furthermore, starting each sequence with an initialization laser, followed by a $\pi$ pulse on the electron spin and a repump laser, allows us to extract the initialization fidelity from
\begin{align*}
F_\mathrm{I,e} = a_1/(a_1 + n_\mathrm{ss,1}),    
\end{align*}
where we neglected background counts, as such potentially underestimating $F_\mathrm{I,e}$.

\newpage
\subsection*{Supplementary Note 3: Photoluminescence excitation spectroscopy.}

\begin{suppfigure}[h]
	\centering
	\includegraphics[width=1\linewidth]{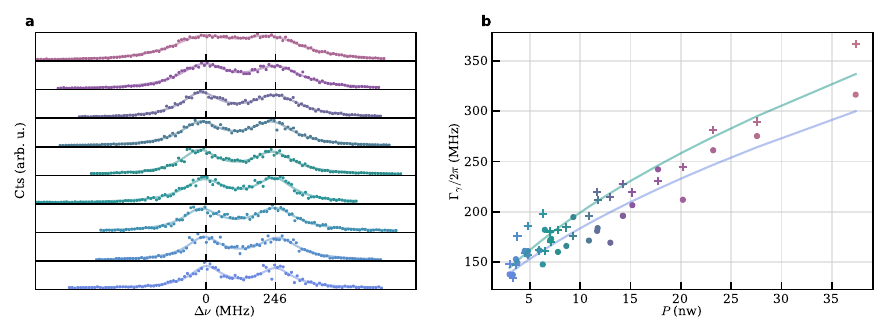}
	\caption{\textbf{Power-dependent photoluminescence excitation measurements (PLE).} 
		\textbf{a} Scanning the laser with frequency $\nu_\mathrm{L}$ and a relative detuning $\Delta \nu = \nu_\mathrm{L} - \nu_\downarrow$ to the $\downarrow$ transition, probes the resonance frequency of the two spin-cycling optical transitions, $\nu_\uparrow$ and $\nu_\downarrow$, while continuously driving the electron spin with a resonant low-power microwave. 
		Solid lines represent fits to the sum of two Lorentzians. 
		\textbf{b} Optical linewidth $\Gamma_\gamma/2\pi$ of left (dot) and right (plus) transition as a function of laser power $P$. Solid line is an extrapolating fit to zero power of the form $\Gamma_{\gamma, 0}/2\pi\sqrt{1+s}$. Colours indicate data from \textbf{a}.}
	\label{fig:PLE}
\end{suppfigure}
We characterize the optical properties by scanning a resonant laser with different excitation powers over the two spin-dependent optical dipoles, see Supplementary Fig.\ref{fig:PLE}a. 
To prevent optical pumping during excitation, we continuously drive the electron spin with a resonant low-power microwave. 
From the collected fluorescence, we then fit two Lorentzians and extract the power-dependent linewidth $\Gamma/2\pi$, which in turn is fit with a typical saturation law with saturation parameter $s=P/P_\mathrm{sat}$
\begin{align}
\Gamma_\gamma(s)  = \Gamma_{0, \gamma} \sqrt{1+s}    
\end{align}
which yields a zero-power optical linewidth of $\Gamma_{0, \gamma}/2\pi = \LinewidthLeft / \LinewidthRight$  for the left/right optical transition, respectively, see Supplementary Fig.\ref{fig:PLE}b. 
This is close to the Fourier limit, $T_\mathrm{1,\gamma}^{-1}/2\pi=\FourierLimit$, which we extracted from the fluorescence decays of the optical control measurements, see Supplementary Fig.\ref{fig:optlifetime}, and thus indicates low inhomogeneous broadening.
Additionally, averaging the splitting between both transitions over all scans yields $\Delta_\mathrm{ss} / 2\pi= \langle  \nu_\uparrow - \nu_\downarrow \rangle = \SplittingAverage$.
\newpage

\subsection*{Supplementary Note 4: Spin-initialization rate and fidelity.}

\begin{suppfigure}[h]
	\centering
	\includegraphics[width=1\linewidth]{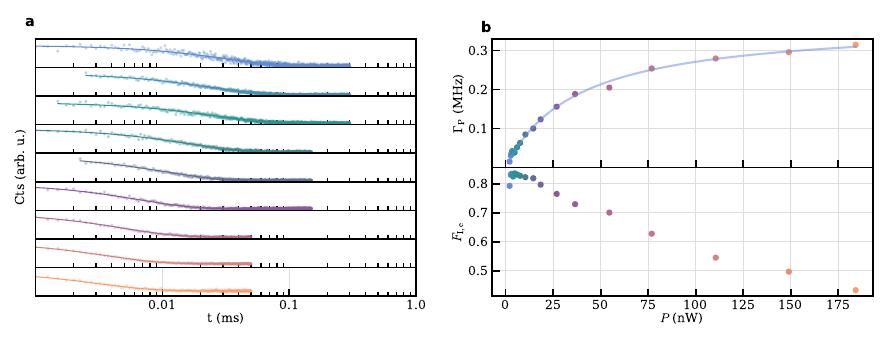}
	\caption{\textbf{Spin-initialization rate and fidelity.} 
		\textbf{a} Power-dependent fluorescence decays due to spin pumping following an initialization step and $\pi$ pulse. 
		\textbf{b} Extracted polarization rate $\Gamma_\mathrm{P}$ and fidelity $F_\mathrm{I, e}$ the upper and lower panel, respectively, as a function of excitation power $P$. Colours indicate data from \textbf{a}. Solid line is a fit to \eqref{eq:polrate}.}
	\label{fig:spinpumping}
\end{suppfigure}
We probe the initialization-fidelity with two resonant laser pulses and with a microwave $\pi$ pulse in-between. The first pulse pumps the spin into its steady dark-state, i.e. the opposite spin state, whereas the second laser pulse then probes the fluorescence after a $\pi$ pulse which inverted the population, see Supplementary Fig.\ref{fig:spinpumping}a. 
Increasing the laser power starts to also depolarize the target spin state again by off-resonantly driving the other optical dipole in the spectral vicinity, see $\Delta_\mathrm{ss}$ from Supplementary Fig.\ref{fig:PLE}. 
We fit an exponential decay to the data and extract the respective polarization rate and fidelity plotted in upper and lower panel of Supplementary Fig.\ref{fig:spinpumping}b, respectively. The polarization rate follows \cite{rosenthal2024single}
\begin{align}\label{eq:polrate}
\Gamma_\mathrm{P} = \frac{\Gamma_0}{2} \frac{1}{\eta} \frac{s}{1+s}, \qquad s=P/P_\mathrm{sat}.
\end{align}  
Fitting the data results in a cyclicity $\eta$ of $\Cyclicity $, see \eqref{eq:cyclcity}. $\Gamma_0 = \FourierLimit$ is the Fourier-limited linewidth, see Supplementary Fig.\ref{fig:optlifetime}. 
We reach a maximum initialization fidelity of $F_\mathrm{I,e}\approx 0.84$, which is limited by remaining spin-mixing and spin-relaxation due to a misaligned magnetic field, as well as off-resonant pumping of the other optical dipole.

\newpage

\subsection*{Supplementary Note 5: Randomized benchmarking.}
\begin{suppfigure}[h]
	\centering
	\includegraphics[width=0.5\linewidth]{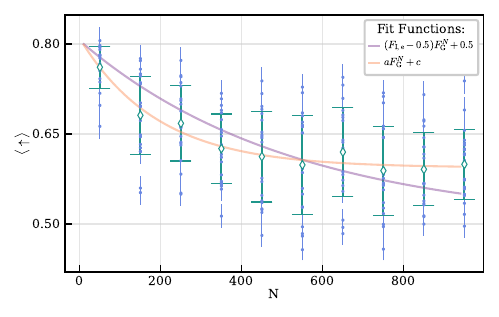}
	\caption{\textbf{Randomized benchmarking.}
		Estimation of single qubit gate fidelity from a set of randomized benchmarking experiments. 
		Blue data points represent the final readout of the sequence. 
		Solid green lines and diamonds represent one standard deviation and mean, respectively. 
	}
	\label{fig:SI21}
\end{suppfigure}
We estimate our single-qubit gate fidelity by performing randomized benchmarking with a Rabi frequency of $\Omega_\mathrm{R}/2\pi = \SI{8.878}{\mega\hertz}$, which is a typical frequency throughout this work. 
To this end, we apply sequences $U_\mathrm{RB}=\Pi_{i=1}^{N} C_i$ consisting of $N$ randomized Clifford operations $C_i$ from the set $\{ \pm X/2, \pm X,  \pm Y/2, \pm Y \}$. 
For each sequence we calculate a final element, which undoes the operation and brings the spin into the opposite state of initialization, where it is then read out. 
Finally, we repeat the randomization $20$ times for each $N$ to collect statistics and then average the results. 
From a fit of the form $(F_\mathrm{I,e}-0.5)\cdot F_\mathrm{G}^N + 0.5$ to the mean values, where we have fixed the amplitude $(F_\mathrm{I,e}-0.5)$ and offset $0.5$, we extract a single-qubit gate fidelity of $F_\mathrm{G}=\FRandBench$ \cite{guo2023microwave, rosenthal2023microwave}. 
We additionally included a fit of the form $a F_\mathrm{G}^N + c$ which fits the data more accurately, resulting in $F_\mathrm{G} = \FRandBenchTwo$. However, from this fit we obtain an offset different from an expected statistical mixture of $0.5$. This can be explained with too few statistics, i.e. more than 20 randomization runs for each $N$ is required or with an erroneous inversion element at the end of each sequence. Nevertheless, we are not limited by either gate fidelity.
\newpage
\subsection*{Supplementary Note 6: Hartmann-Hahn cross relaxation.}

\begin{suppfigure}[h]
	\centering
	\includegraphics[width=1\linewidth]{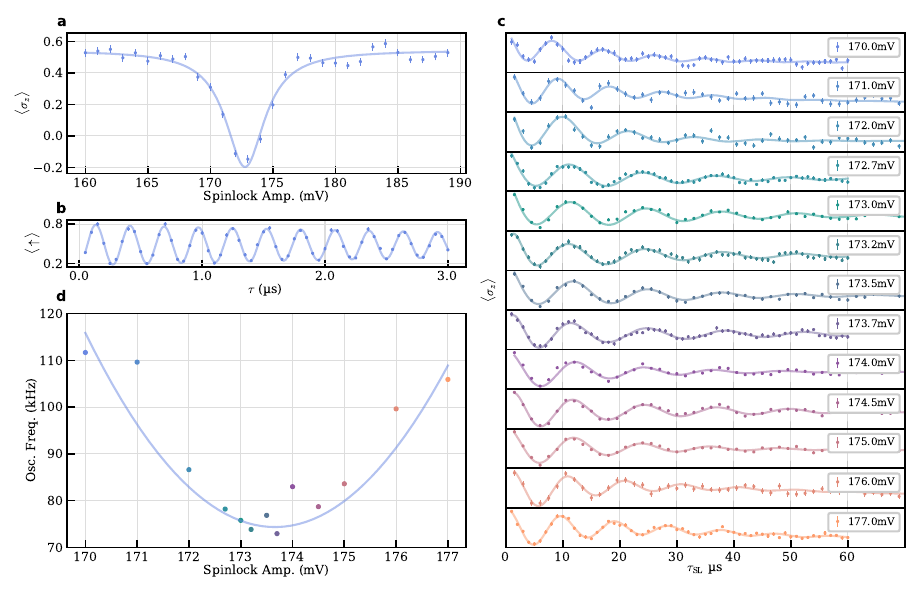}
	\caption{\textbf{Power-dependent Hartmann-Hahn.}
		\textbf{a} Sweeping the amplitude of a 90° out of phase microwave signal with respect to an initial $\pi/2$ pulse at constant $\tau$ reveals loss of coherence when the driving becomes resonant with the spin bath. 
		Solid line is a Lorentzian fit. 
		\textbf{b} Electron Rabi oscillations with a microwave amplitude corresponding to the center of the Hartmann-Hahn resonance from \textbf{a}. 
		\textbf{c} Varying the spin-lock time $\tau_\mathrm{SL}$ reveals coherent oscillations. 
		Solid lines are fits to a sum of two exponentially damped sines. 
		\textbf{d} Dominant frequency component of fits from \textbf{c}, solid line is a parabola fit to the data. }
	\label{fig:SI3}
\end{suppfigure}
We try to independently estimate $A_\perp$ by probing electron-nuclear cross relaxation by locking the electron spin with a Rabi frequency $\Omega_\mathrm{SL}$ close to the nuclear spins Larmor frequency. 
This is achieved by first initializing the electron spin, applying a $\pi/2$ pulse to rotate the spin onto the equator of the Bloch sphere and then driving the spin 90° out of phase relative to the first $\pi/2$ pulse for a duration $\tau_\mathrm{SL}$, i.e. driving it along its eigenstate. 
Finally, we measure the remaining coherence by performing another $\pi/2$ followed by a spin-readout \cite{metsch2019initialization}.\\
In Supplementary Fig.\ref{fig:SI3}a the spin-lock amplitude of the 90° out of phase microwave signal is varied with a constant length of $\tau_\mathrm{SL}$. 
If the corresponding $\Omega_\mathrm{SL}$ gets into resonance with the Larmor frequency $\omega_\mathrm{L}$ of the $^{13}\mathrm{C}$ nuclear spin bath, the electron spin's population is coherently transferred onto the nuclear spins, thereby losing coherence, clearly visible in Supplementary Fig.\ref{fig:SI3}a. 
We used $\tau_\mathrm{SL} = \SI{7.246}{\micro\second} \approx 2\pi/A_\perp$, where we expected a maximum transfer.\\
With the resonant spin-lock amplitude of Supplementary Fig.\ref{fig:SI3}a, we perform a Rabi measurement resulting in $\Omega_\mathrm{R}/2\pi = \RabiHH$, see Supplementary Fig.\ref{fig:SI3}b, from which we infer $\omega_{\mathrm{L,n}}/2\pi=\Omega_\mathrm{R}/2\pi=\RabiHH$, close to the Larmor frequency determined from the dynamical decoupling resonances, $\omega_{\mathrm{L,n}}/2\pi=\wLXYfourtytwo$. 
We attribute the slight mismatch to the amplitude granularity of our arbitrary waveform generator (AWG).\\
Next, we fix $\Omega_\mathrm{SL}$ at various amplitudes around the resonance and vary the spin-lock time $\tau_\mathrm{SL}$. 
The data is shown in Supplementary Fig.\ref{fig:SI3}c where different oscillations in the form of cross-relaxation of the electron spin's population are visible, which are a direct result of coherent coupling to nearby, individual nuclear spins. 
The signal can not be described well by a single, exponentially damped sine and is therefore fitted with the sum of two sines, indicating coupling to more than one $^{13}C$ spin. \\
Supplementary Fig.\ref{fig:SI3}d displays the extracted prominent frequency components of Supplementary Fig.\ref{fig:SI3}c as a function of spin-lock amplitude, revealing an increase in cross-relaxation frequency when $\Omega_\mathrm{SL}$ is detuned from the Hartmann-Hahn condition, $\Omega_\mathrm{SL} = \omega_\mathrm{L,n}$. 
Therefore, we fit the data with a parabola leading to a minimum of  $A_\perp/2\pi \approx 2\cdot \AperpHHParabola$, slightly higher compared to the fits of XY-DD sequences in the main text, which we attribute again to the amplitude granularity of the AWG.\\
The data from Supplementary Fig.\ref{fig:SI3}a and d have been measured at different days. 
A slightly different microwave field at the sample can explain the small discrepancy in resonant spin-lock amplitude, i.e. respective minima in Supplementary Fig.\ref{fig:SI3}a and d.
\newpage

\subsection*{Supplementary Note 7: Temperature-dependent rates.}
\begin{suppfigure}[h]
	\centering
	\includegraphics[width=1\linewidth]{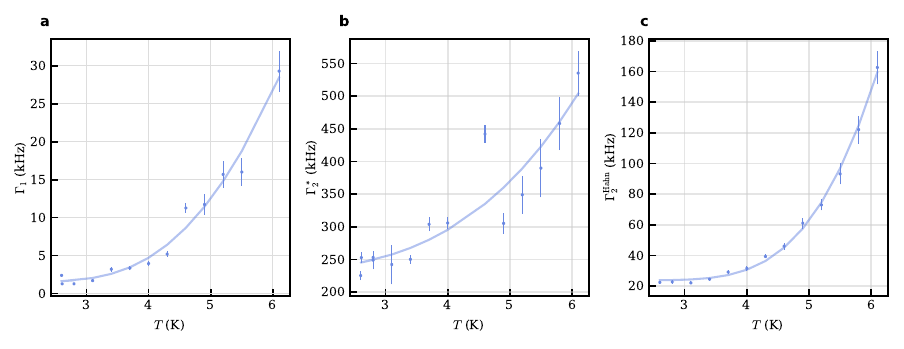}
	\caption{\textbf{Temperature dependent rates.}  
		\textbf{a} Electron spin relaxation $\Gamma_\mathrm{1}$, \textbf{b} dephasing $\Gamma^*_\mathrm{2}$ and \textbf{c} decoherence $\Gamma^{\mathrm{Hahn}}_2$ rates. 
		We fit the data in \textbf{a} and \textbf{c} with a two-phonon Orbach process. Data in \textbf{b} is fit with a power law of the form $\Gamma^*_\mathrm{2} = a T^b +c$.
	}
	\label{fig:TempRates}
\end{suppfigure}
In Supplementary Fig.\ref{fig:TempRates} we show temperature-dependant measurements of the electron spin relaxation $\Gamma_\mathrm{1}$, dephasing $\Gamma^*_\mathrm{2}$ and decoherence $\Gamma^{\mathrm{Hahn}}_2$ rates. 
We fit the rates $\Gamma_\mathrm{1}$ and $\Gamma^{\mathrm{Hahn}}_2$ with a resonant two-phonon Orbach process and a temperature-independent offset $\Gamma_\mathrm{0}$ \cite{guo2023microwave, orbach1961spin}. 
Following \cite{guo2023microwave}, we also introduce a phenomenological parameter $\alpha$ leading to an effective temperature $T_\mathrm{eff} = \alpha T$
\begin{align*}
\Gamma(T) = \Gamma_\mathrm{0} + \Gamma_\mathrm{Orbach}(T) = \Gamma_\mathrm{0} + \frac{a\cdot \Delta_\mathrm{gs}^3}{\exp\left(\frac{\hbar\Delta_\mathrm{gs}}{k_\mathrm{B} T_\mathrm{eff}}\right)-1} \; .
\end{align*}
This leads to $\alpha_1 = \alphaone$ and $\alpha_2^\mathrm{Hahn} = \alphatwo$. 
Using the latter, we can conservatively estimate the sample's effective temperature at $T\approx \SI{4}{\kelvin}$ and higher.\\
We measure the temperature inside the flow-cryostat at the helium exchanger, away from the sample, which is mounted on a copper coldfinger. 
Thus, we expect the temperature at the sample to be higher, due to the limited thermal conductivity through many interfaces and induced heat-load by our room temperature high-NA objective. 
Previous studies with the same experimental setup estimated local defect temperatures in the range of 4-8K \cite{klotz2022prolonged}, inline with our effective sample temperature. 
In contrast to recent reports for tin vacancy centers \cite{rosenthal2023microwave, guo2023microwave} we observe an increase of $\Gamma^*_\mathrm{2}\propto T^{\bttwostar}$, where $\bttwostar$ hints at higher-order phonon processes.
\newpage
\subsection*{Supplementary Note 8: Nuclear spin initialization and hyperfine parameters.}
\begin{suppfigure}[h]
	\centering
	\includegraphics[width=1\linewidth]{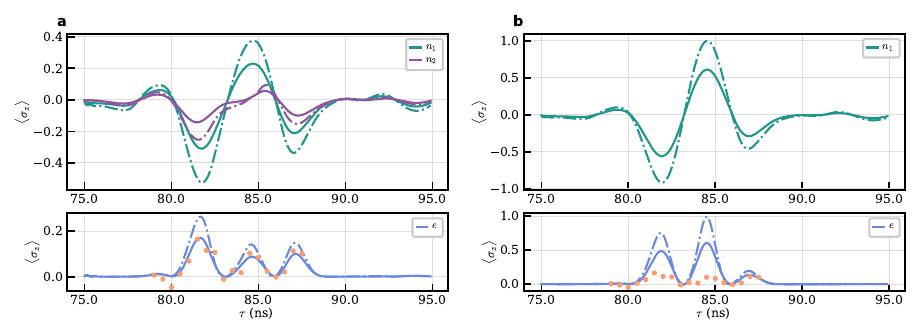}
	\caption{\textbf{Nuclear spin initialization and hyperfine parameters.}
		Simulation of the nuclear spin initialization gate $U_\mathrm{I}$ from Fig.3b of the main article, adapted from \cite{sipahigil2016integrated, taminiau2014universal}. 
		Solid lines are simulations with the experimentally determined electron spin initialization of $F_\mathrm{I,e} = 0.84$ and $N_\mathrm{I} = 42$. 
		Dash-dotted lines are ideal cases with $F_\mathrm{I,e}=1$ for comparison. 
		Parameters can be extracted from Tab.\ref{tab:DDparams}. The time for a $\pi$ pulse is $T_\mathrm{\pi} = \SI{55.715}{\nano\second}$. \\
		\textbf{a} Expectation values of $\sigma_z$ of the target nuclear spin $n_1$ with a parasitic auxiliary nuclear spin $n_2$.
		\textbf{b} Same as \textbf{a} but with only the target nuclear spin $n_1$.
		Respective lower panels indicate the measured electron spin populations (dots) and simulated populations (solid and dash-dotted lines) after a reversed $U_\mathrm{I}$, which probes the nuclear spin, see measurement and description in the main text. 
	}
	\label{fig:SINucInit}
\end{suppfigure}
Since for spin $1/2$ systems the resonance condition of weakly coupled nuclear spins is proportional to $\left(A_\perp/\omega_\mathrm{L,n}\right)^2$ \cite{nguyen2019quantum, zahedian2024blueprint}, long inter-pulse spacings $\tau$ during dynamical decoupling are necessary to resolve individual spins.\\
However in this work, the electron spin coherence time was limited and we could not resolve individual nuclear spins. 
Thus, we used the first resonance at $\tau = T_\mathrm{L,n}/2 - T_\mathrm{\pi}$, where $T_\mathrm{L,n}$ is the nuclear Larmor period, to initialize and coherently rotate the nuclear spin with gates adapted from \cite{sipahigil2016integrated, taminiau2014universal}. 
This enables fast gate times due to the short inter-pulse spacing. 
However, the overlap of the target nuclear spin's resonance with the nuclear spin bath lead to initialization infidelity such that we included an auxiliary effective nuclear spin in our simulations, which accounts for parasitic effects. \\
In Supplementary Fig.\ref{fig:SINucInit} the simulated $\sigma_z$ of the involved spins (see Methods for model Hamiltonian) after performing the initialization gate $U_\mathrm{I}(\tau,N_\mathrm{I})$ with $N_\mathrm{I}=42$ $\pi$ pulses are depicted in Supplementary Fig.\ref{fig:SINucInit}a and Supplementary Fig.\ref{fig:SINucInit}b with and without the additional nuclear spin, respectively. 
The solid/dash-dotted lines represent the simulation with the experimentally determined/ideal electron initialization of $F_\mathrm{I,e} = \SI{0.805946119102144 \pm 0.00426983790769161}{}$/$1$. 
Comparing the measurement outcomes in each lower panel of Supplementary Fig.\ref{fig:SINucInit}a and b with the corresponding simulation (solid line), we can see that only taking a single nuclear spin into consideration overestimates the measurement outcome by roughly a factor of $2$, whereas two nuclear spins already are in good agreement with the data, see Supplementary Fig.\ref{fig:SINucInit}. \\
In the top panel of Supplementary Fig.\ref{fig:SINucInit}a, we can see that besides our target nuclear spin, the additional nuclear spin also gets population transferred, effectively lowering the target initialization fidelity to $F_\mathrm{I, n_1} = 0.647$.\\
In Tab.\ref{tab:DDparams} we list the fitted free nuclear spin parameters for both nuclear spins from the main article's Fig.3 with one standard deviation as error. 
One set of parameters could not describe the whole data set, which we ascribe to the sensitivity of the experiment to external fluctuations and simplicity of the model only containing two nuclear spins. 
However, it is worth noting that the perpendicular hyperfine parameters of the target nuclear spin are equal within their respective errors.
The target nuclear spin's parallel hyperfine parameter $A_{\parallel,1}$ has been fixed to \Aparallel, whereas $A_\mathrm{\parallel, 2}$ reached the upper bound of $\SI{50}{\kilo\hertz}$, which we roughly estimated from $(T_\mathrm{2,e}^*)^{-1} /4 \approx \SI{50}{\kilo\hertz}$.
\begin{table}[h]
	\centering
	\begin{tabular}{c|c|c|c|c}
		Fig. 3 & Sequence & $\omega_\mathrm{L,n}/2\pi$ & $A_{\perp,1}/2\pi$ & $A_{\perp,2}/2\pi $ 
		\\ \hline
		a & XY-24 &\wLXYtwentyfour& \AperpOneXYtwentyfour & \AperpTwoXYtwentyfour
		\\
		& XY-32 &\wLXYthritytwo& \AperpOneXYthritytwo & \AperpTwoXYthritytwo 
		\\
		& XY-42 &\wLXYfourtytwo& \AperpOneXYfourtytwo & \AperpTwoXYfourtytwo 
		\\
		& XY-64 &\wLXYsixtyfour& \AperpOneXYsixtyfour & \AperpTwoXYsixtyfour 
		\\ \hline
		b & & from XY-42  & \AperpOneXYInitRead & \AperpTwoXYInitRead\\ \hline
		d & Cond. Rot. & \wLXYRcond & \AperpOneXYRcond & \AperpTwoXYRcond\\
		& Uncond. Rot. & \wLXYRuncond & \AperpOneXYRuncond & \AperpTwoXYRuncond \\
	\end{tabular}
	\caption{Fitted nuclear spin parameters from the main article's Fig.3.} 
	\label{tab:DDparams}
\end{table}
\newpage

\subsection*{Supplementary Note 9: Nuclear spin dephasing and decoherence.}

\begin{suppfigure}[h]
	\centering
	\includegraphics[width=.5\linewidth]{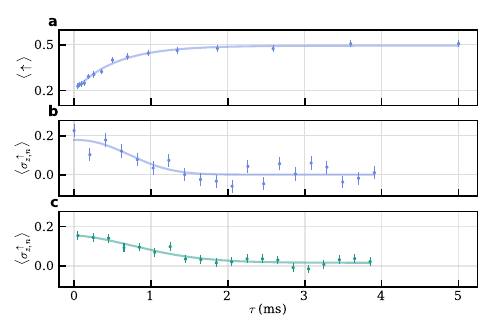}
	\caption{\textbf{Nuclear spin dephasing and decoherence.} \textbf{a} Electron spin relaxation time. Solid line is an exponential recovery fit to the data. 
		\textbf{b} Nuclear Ramsey measurement. 
		Solid line is a fit with a stretched-exponential decay $a\exp-(t/T^{*}_\mathrm{2,n})^\beta + c$.
		\textbf{c}  Nuclear spin Hahn echo. 
		Solid line is a stretched-exponential decay.}
	\label{fig:NucCoh}
\end{suppfigure}
At $T=\SI{2.6}{\kelvin}$ base temperature, we measure the spin-relaxation time through a pump-probe experiment, see Supplementary Fig.\ref{fig:NucCoh}a.
To this end, we first initialize the electron spin out of thermal equilibrium and then increase a free evolution time $\tau$ before reading it out again. 
This effectively probes the electron's population decay back in to equilibrium which results in a relaxation time of $T_{1, \mathrm{e}} = \TimeOne$.
After determining the number $N_{\pi/2}$ of $\pi$ pulses at $\tau_\mathrm{rot} = T_\mathrm{L}/2 - T_\pi$, which rotates the target nuclear spin by $\pi/2$ during XY-$N$ dynamical decoupling, we performed a Ramsey interference experiment identical to the one shown in Fig.3e in the main text. 
In contrast to Fig.3e from the main text, we are not interested in the electron spin's dependent precession frequency which is why we coarsely chose $\tau$ in the measurement shown in Supplementary Fig.\ref{fig:NucCoh}b and c. 
From an stretched-exponential fit to the data, we extract a nuclear spin dephasing time $T_2^* = \TimeTwoStarNuclear$ with a stretching factor $\beta = \BetaNTimeTwoStar$. 
Adding a single refocusing $\pi$ pulse on the nuclear spin extends the coherence time to $T_2 = \TimeTwoNuclear$  with a stretching factor $\beta = \BetaNTimeTwo$, which we obtained similarly from Supplementary Fig.\ref{fig:NucCoh}c. \\
The coherence time $T_2$ is well within the limits of $2T_\mathrm{1,e} = \SI{1.070\pm 0.104}{\milli\second}$ suggesting that spin-lattice relaxation is the limiting factor, preventing longer coherence times.
\newpage
\subsection*{Supplementary Note 10: Low-power Rabi measurements.}

\begin{suppfigure}[h]
	\centering
	\includegraphics[width=1\linewidth]{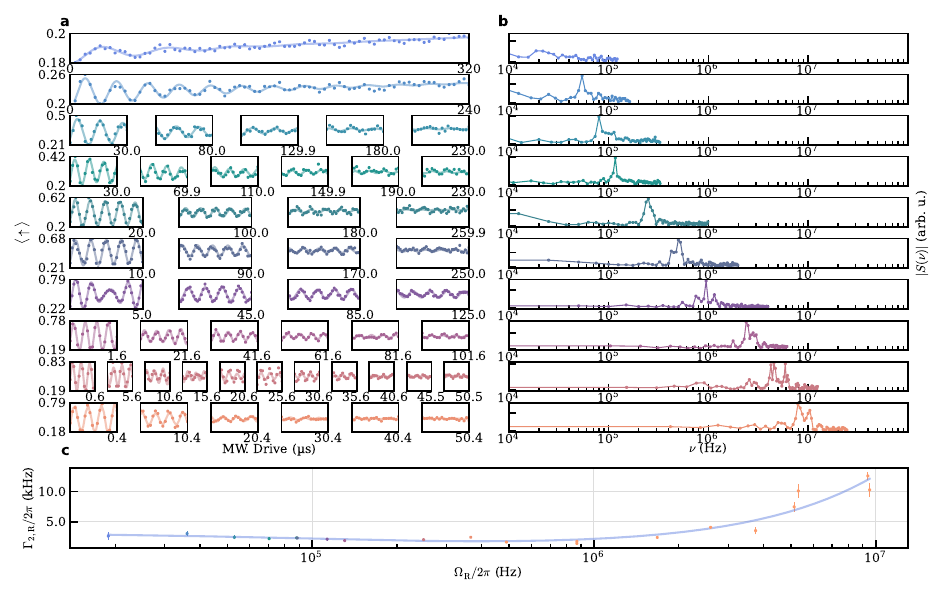}
	\caption{\textbf{Extended Rabi measurements.} 
		\textbf{a} Rabi measurements for various driving strengths, resonant with one of the hyperfine split transitions of the nuclear spin with $A_\parallel / 2 \pi = \Aparallel$.
		\textbf{b} Fourier transform of \textbf{a}.
		\textbf{c} Decoherence rate $\Gamma_\mathrm{2,R}/2\pi$ as a function of Rabi frequency $\Omega_\mathrm{R}/2\pi$. Solid line is a phenomenological fit of the form $a/\Omega_\mathrm{R} + b\Omega_\mathrm{R} + c$.
	}
	\label{fig:contDD}
\end{suppfigure}
We performed power-dependent Rabi measurements in order to explore the coherence limits while continuously decoupling the electron spin. 
Supplementary Fig.\ref{fig:contDD}a shows several such measurements together with its Fourier transform in Supplementary Fig.\ref{fig:contDD}b. 
The smallest Rabi frequency we resolved was $\Omega_\mathrm{R} = \LowRabi$, indicating the potential for resolving nuclear spins with $A_\parallel\ll\Aparallel$. 
We obtained a minimum in decoherence rate of $\Gamma_\mathrm{2,R}=\LowGammaTwoR$, slightly higher than what we have achieved during the spin-locking measurement in Fig.2b of the main text.\\
In Supplementary Fig.\ref{fig:contDD}c the decoherence rate is decreasing with $1/\Omega_\mathrm{R}$ until $\Omega_\mathrm{R} \approx \SI{500}{\kilo\hertz}$ which we attribute to the increasing dressed-states' energy barrier $\Omega_\mathrm{R}$ which effectively suppresses coupling to noise.
Moreover, above the threshold increasing driving-induced heating might result in an increasing decoherence rate with $\Omega_\mathrm{R}$. 
The fact that we did not reach the lower decoherence rate limit $\Gamma_\mathrm{2,R} = 1/2T_\mathrm{1,e} \approx \SI{150}{\hertz}$ is attributed to a driving-induced amplitude noise \cite{senkalla2024germanium}.
\newpage
\subsection*{Supplementary Note 11: Nuclear spin single-shot readout.}
\begin{suppfigure}[h]
	\centering
	\includegraphics[width=1\linewidth]{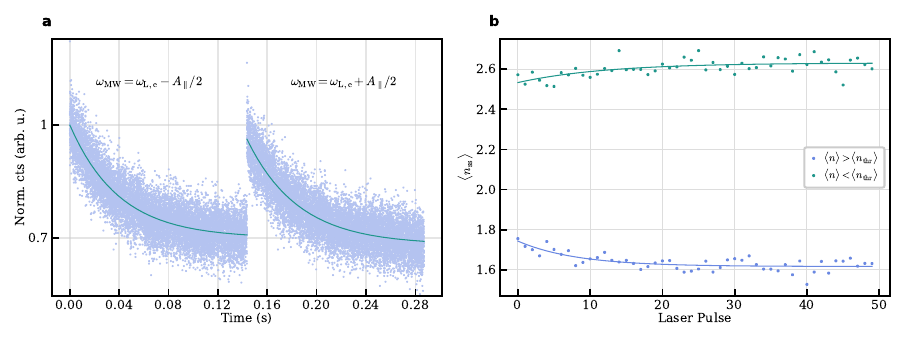}
	\caption{\textbf{Single-shot readout.}
		\textbf{a} Summed counts (blue data) during each block of optical electron spin polarization for $T_\mathrm{p} = \SI{10}{\micro\second}$ followed by a $\mathrm{C_nNOT_e}$ gate. 
		Counts are normalized to the maximum of the fit. Solid green line is a fit to an exponential decay.
		After \SI{140}{\milli\second} the driving frequency is changed to the opposite nuclear spin resonance.
		\textbf{b} Time-binned, mean-photon number $\langle n_\mathrm{ss}\rangle$ of each electron spin readout laser pulse within the steady state during a Ramsey measurement after a SSR window of $T_\mathrm{SSR}=\SI{3}{\milli\second}$. See Fig4.c and d of the main text. 
		Solid lines are exponential fits.
	}
	\label{fig:SSR}
\end{suppfigure}
We investigate optical pumping of the nuclear spin, shown in Supplementary Fig.\ref{fig:SSR}a by increasing the single-shot readout (SSR) window to $T_\mathrm{SSR} = N_\mathrm{SSR}\cdot(T_\mathrm{p} + T_\mathrm{\mathrm{C_nNOT_e}}) \approx \SI{140}{\milli\second}$, see main text for further details. 
After $T_\mathrm{SSR}$ we change the electron spin's driving frequency to the opposite nuclear spin state resonance to reverse the optical polarization process.
In the experiment shown in Supplementary Fig.\ref{fig:SSR}a we used $N_\mathrm{SSR} = 12000$, $T_\mathrm{p}=\SI{10.5}{\micro\second}$ and $T_\mathrm{\mathrm{C_nNOT_e}} = \SI{1.433}{\micro\second}$. 
Fitting exponential decays, $a\exp(-t/T_\mathrm{p, n}) +c$ to the summed counts results in a optical nuclear spin polarization timescales of $T_\mathrm{p, n} = \PumptimeNuclear$ and $\PumptimeNuclearRev$ with a contrast of $c/(a+c) =\PumpContrastNuclear$ and $\PumpContrastNuclearRev$ for both nuclear spins, respectively.
From the obtained $T_\mathrm{p, n}$ we can conclude that we are not polarizing the nuclear spin significantly, as long as $T_\mathrm{SSR}\ll T_\mathrm{p, n}$. 
However, in the main text we used a $T_\mathrm{SSR} = \SI{3}{\milli\second}$ such that we estimate an on-set of polarization and thereby loss of nuclear spin initialization fidelity by $1-\exp (-\SI{3}{\milli\second} / \SI{41.62}{\milli\second}) \approx 0.07$ in the Ramsey measurement, inline with the extracted initialization fidelity $F'_\mathrm{I,n_b}$ from the mean-photon statistics of the main text.\\
Furthermore, we observed an exponential increase/decrease in the steady-state counts within each electron spin readout laser pulse after a $T_\mathrm{SSR} = \SI{3}{\milli\second}$ SSR and respective classification into a bright and dark nuclear spin state, see Supplementary Fig.\ref{fig:SSR}b. We attribute this effect to a bright/dark SiV state, which is being pumped dark/bright during consecutive laser pulses. 
This was not visible in other measurements, where no post-selection was done. 
Thus, we conclude that we also classify and post-select the SiV's bright/dark state, independent of the nuclear spin state. 
Further studies have to be done in order to properly discriminate these two effects which should improve the nuclear spin's initialization fidelity.
In order to correct the measured Ramsey interference data, we normalize each extracted amplitude $a_i$, see Supplementary Fig.\ref{fig:pulsefidelity}, by the fitted relative increase/decrease in steady-state counts.
\newpage
\subsection*{Supplementary Note 12: Pulse-width and optical lifetime.}
\begin{suppfigure}[h]
	\centering
	\includegraphics[width=1\linewidth]{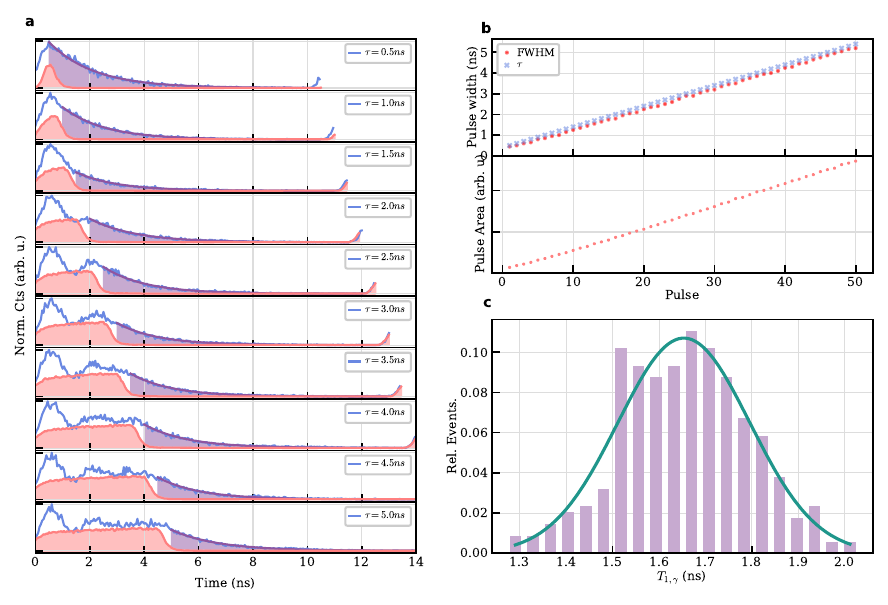}
	\caption{\textbf{Pulse-width and optical lifetime.}
		\textbf{a} Time resolved fluorescence signal (blue) showing oscillations during continuous driving of the SiV's optical dipoles. Purple data is an exponential fit to the fluorescence decay, proportional to the excited state population. Red data is the excitation laser reference signal. Data set is from Fig.5a of the main text, where every 5th point of the modulation time $\tau$ is shown.
		\textbf{b} Top-panel: Full-width-half-maximum (FWHM) of excitation laser pulses (red) and modulation time (blue). Lower-panel: Excitation pulse area as extracted from the laser reference signal.
		\textbf{c} Distribution of fluorescence decay constants $T_\mathrm{1,\gamma}$ from the complete collection of Rabi measurements of Fig.5a from the main text. Green solid line is a fit to a Gaussian distribution.
	}
	\label{fig:optlifetime}
\end{suppfigure}
Supplementary Fig.\ref{fig:optlifetime}a shows an extended dataset for the Rabi measurements with a modulation amplitude of $\SI{100}{\milli\volt}$, see top panel of Fig.5a of the main text. 
Here, the time-tagged fluorescence is shown for every 5th modulation time $\tau$. \\
We extract the rising edge of the fluorescence signal and add $\tau$ to it to determine the start of the fluorescence decay. 
A subsequent exponential fit to the remainder of the fluorescence, see purple line in Supplementary Fig.\ref{fig:optlifetime}a, allows us to obtain the amplitude, proportional to the excited state population, as well as decay constant $T_\mathrm{1, \gamma}$, which is the dipole's optical lifetime. \\
Supplementary Fig.\ref{fig:optlifetime}b shows the excitation laser's pulse width (FWHM) and area in the top and lower panel, respectively, closely following the modulation time $\tau$'s linear increase. This verifies the quality of the optical pulses. The difference of FWHM and $\tau$ is attributed to the EOM's finite risetime.\\
Supplementary Fig.\ref{fig:optlifetime}c shows statistics of the optical lifetime $T_\mathrm{1, \gamma}$. A Gaussian distribution fit to the data results in a mean $T_\mathrm{1,\gamma}$ of $ \TimeOneOptical$, where the error is the Gaussian's standard deviation.

\newpage
\printbibliography
	
\end{document}